\documentclass[article,onecolumn]{revtex4}
\usepackage{makeidx}
\usepackage{amssymb}
\usepackage{amsmath}
\usepackage{graphicx}
\usepackage{dcolumn}
\usepackage{bm}
\usepackage[center]{subfigure}
\usepackage{color}
\usepackage[bf, small]{titlesec}
\usepackage[symbol]{footmisc}

\renewcommand\thesection{\arabic{section}}
\renewcommand\thesubsection{\arabic{section}.\arabic{subsection}}
\titleformat{\section}{\bf\normalsize}{\thesection.\,}{0.5pt}{}
\titlespacing{\section}{0cm}{12pt}{12pt}
\titleformat{\subsection}{\bf\small}{\thesubsection.\enspace}{0.5pt}{}
\titlespacing{\subsection}{0cm}{12pt}{12pt}
\providecommand{\keywords}[1]{\small\textbf{\textit{Keywords---}} #1}

\begin{document}

\title{Stabilization of one-dimensional Townes solitons by spin-orbit
coupling in a dual-core system}
\author{Elad Shamriz$^{a}$, Zhaopin Chen$^{a}$\thanks{%
Corresponding author.}}
\email{viskolczp@gmail.com}
\author{Boris A. Malomed$^{a,b}$}
\affiliation{$^{a}$Department of Physical Electronics, School of Electrical Engineering,
Faculty of Engineering, and Center for Light-Matter Interaction, Tel Aviv
University, P.O.B. 39040, Ramat Aviv, Tel Aviv, Israel\\
$^{b}$Instituto de Alta Investigaci\'{o}n, Universidad de Tarapac\'{a},
Casilla 7D, Arica, Chile}

\begin{abstract}
It was recently demonstrated that 2D Townes solitons (TSs) in two-component
systems with cubic self-focusing, which are normally made unstable by the
critical collapse, can be stabilized by linear spin-orbit coupling (SOC), in
Bose-Einstein condensates and optics alike. We demonstrate that 1D TSs,
realized as optical spatial solitons in a planar dual-core waveguide with
dominant quintic self-focusing, may be stabilized by SOC-like terms emulated
by obliquity of the coupling between cores of the waveguide. Thus, SOC
offers a universal mechanism for the stabilization of (quasi-) TSs. A
combination of systematic numerical considerations and analytical
approximations identifies a vast stability area for skew-symmetric solitons
in the system's main (semi-infinite) and annex (finite) bandgaps. Tilted
(\textquotedblleft moving") solitons are unstable, spontaneously evolving
into robust breathers. For broad solitons, diffraction, represented by
second derivatives in the system, may be neglected, leading to a simplified
model with a finite bandgap. It is populated by skew-antisymmetric gap
solitons, which are nearly stable close to the gap's bottom.
\end{abstract}

\maketitle

\section{Introduction}

A well-known problem in studies of solitons (self-trapped modes in nonlinear
dispersive/diffractive media) is that, if the focusing nonlinearity is too
strong and/or underlying spatial dimension is too high, soliton solutions
are destabilized by the presence of the collapse, i.e., catastrophic
self-compression of wave fields in the same media \cite{Berge,Fibich}. If
the respective $D$-dimensional equation of the nonlinear-Schr\"{o}dinger
(NLS) type for complex field $u$ contains a focusing term $\sim |u|^{2\sigma
}u$ with real $\sigma >0$, the critical collapse (the onset of which
requires a norm of the field exceeding a certain threshold value) takes
place at $\sigma D=2$, and the supercritical collapse (which is possible
with an arbitrarily small norm) occurs at $\sigma D>2$ \cite{Berge,Fibich}.
Thus, the critical nonlinearity strength corresponds to $\sigma =1$ (cubic)
and $\sigma =2$ (quintic) in the 2D and 1D geometries, respectively. A
specific feature of the NLS equation with the critical nonlinearity is the
existence of a degenerate family of \textit{Townes solitons }(TSs), first
found as 2D numerical solutions of the cubic NLS equation \cite{Townes}, and
identified as a 1D analytical solution of the quintic equation \cite%
{1DT1,1DT2}. The families are degenerate in the sense that all the TSs share
a common value of the integral norm, due to the fact that the NLS equation
with the critical nonlinearity features a conformal invariance, allowing one
to transform different TSs into each other, keeping the norm unaltered. The
TSs are unstable solutions, which separate decaying and collapsing ones \cite%
{Berge,Fibich}). In terms of the evolution of small perturbations, the TS
instability is represented by zero eigenvalues, i.e., the instability is
subexponential (but, nevertheless, quite tangible) \cite{Fibich}.

Because the cubic focusing is a ubiquitous nonlinearity in optics (the Kerr
effect) \cite{KA}, plasmas (the nonlinearity of Langmuir waves \cite%
{Robinson}), atomic Bose-Einstein condensates, BECs (the cubic term in the
Gross-Pitaevskii equation induced by attractive inter-atomic collisions \cite%
{Pit}), etc., and all these settings naturally occur in the 2D form, a
challenging problem is to modify the respective models by including
additional physically relevant terms which may stabilize 2D solitons. In the
course of theoretical and experimental work, various solutions of this
problem were elaborated, as summarized in early reviews \cite{2005,2008} and
updated ones \cite{viewpoint,2016,2019}. In particular, these are
higher-order defocusing nonlinearities (usually represented by quintic
terms), which compete with the cubic focusing in optical media \cite%
{Michinel,Cid1,Cid2}, and spatially-periodic potentials, induced by optical
lattices in BEC or by photonic-crystal structures in optical waveguides. The
periodic potentials readily stabilize both fundamental and
vorticity-carrying 2D solitons \cite{BBB,Ziad}. More recently, it was
predicted \cite{Petrov,PA,Polish,Santos}, and experimentally demonstrated in
various forms \cite{Leticia1,Leticia2,Inguscio1,Inguscio2,hetero}, that
self-trapped 3D and quasi-2D states in binary BEC with attraction between
its two components can be created in the form of stable \textquotedblleft
quantum droplets". They are filled by a nearly incompressible superfluid,
being stabilized against the collapse by the effective quartic
self-repulsion induced by the Lee-Huang-Yang (LHY)\ effect \cite{LHY}, i.e.,
a contribution of quantum Bogoliubov modes excited around the mean-field
states. It was predicted too that 3D and 2D \textquotedblleft droplets" with
embedded vorticity\ may be stable as well \cite{Barcelona,Raymond}.
Furthermore, it was found theoretically and demonstrated experimentally that
the LHY effect helps to stabilize localized multidimensional states in BEC
with long-range interactions between dipolar atoms \cite%
{Pfau1,Pfau2,Pfau3,Pfau3,Pfau4,Pfau5}. \

The current work on BEC has suggested another possibility to stabilize 2D
and 3D solitons, namely, the use of the (pseudo-) spin-orbit-coupling (SOC),
which was implemented, as a linear effect, in binary condensates \cite%
{Spielman,Goldman,Zhai}. It was demonstrated that, if linear SOC terms,
which mix two BEC\ components via the first spatial derivatives (which
represent the anomalous velocity in the condensate superfluid \cite%
{super1,super2,XiChen}), are added to the usual cubic intra- and
inter-component attraction, they lift the above-mentioned conformal
invariance, and thus allow the coupled NLS equations to create solitons with
the norm falling below the fixed TS value. As a result, these solitons
become stable (immune to the onset of the critical collapse). In particular,
they restore the system's ground state, which is absent when the dynamics is
dominated by the critical collapse \cite{BLi,Evgeny}. Furthermore, 2D
solitons can be stabilized by the quasi-1D SOC, which is applied along a
single direction in the 2D plane \cite{quasi1D}. In the 3D setting, the
supercritical collapse cannot be suppressed by the SOC terms; nevertheless,
they help to make metastable solitons, which stay robust against small
perturbations \cite{HPu}.

While the (pseudo-) SOC terms in BEC emulate genuine SOC, originally
discovered in semiconductors, by mapping the spinor wave functions of
electrons, moving in the ionic lattice, into a two-component mean-field
bosonic wave function of the binary BEC, the latter setting may be, in turn,
emulated by bimodal light propagation in optical waveguides. In particular,
the above-mentioned mechanism stabilizing 2D matter-wave solitons can be
reproduced for 2D spatiotemporal solitons in a dual-core planar waveguide
(coupler) \cite{emulation}. In the coupler model, two components of the
binary BEC are replaced by optical fields in parallel tunnel-coupled cores
with intrinsic Kerr (focusing cubic) nonlinearity in each core, while the
SOC itself is emulated by temporal dispersion of the linear inter-core
coupling. As a result, a family of stable 2D solitons was constructed. A
similar but different optical model was proposed in Ref. \cite{NJP}, which
was also based on the spatiotemporal propagation of light in a dual-core
planar structure with the Kerr nonlinearity, but the SOC terms were emulated
by spatial (rather than temporal) corrections to the linear inter-core
coupling, produced by obliquity of the barrier separating the parallel cores.

The predicted stabilization of the two-component 2D solitons in the cubic
self-focusing media by the optically emulated linear SOC terms suggests to
consider a possibility to stabilize, by means of optically emulated SOC, 1D
solitons of the quasi-TS type in a medium with quintic focusing, which, as
mentioned above, is the critical nonlinearity in 1D (similarly to the cubic
focusing in 2D, it also gives rise to the critical collapse). The objective
of the present work is to introduce such a model (in the full form including
paraxial diffraction, and its reduced diffractionless version), construct
solitons in it, and investigate their stability. It is relevant to mention
that peculiarities of the collapse of 1D solitons in the SOC model with
quintic attraction were recently addressed in Ref. \cite{XiChen}, but a
possibility of stabilization was not considered there.

The model is introduced below in Section 2. As the preliminary stage of the
analysis, spontaneous symmetry breaking (SSB) in the coupler with the
quintic self-focusing in the absence of SOC is considered in Section 3. The
main part of the work is reported in Section 4, \textit{viz}., systematic
numerical results for families of skew-symmetric solitons stabilized by the
(pseudo-) SOC. Gap solitons in the reduced model, which neglects the second
spatial derivatives, are considered in Section 5. The paper is concluded by
Section 6.

\section{The model}

We start by considering the propagation of optical waves in the planar
dual-core guide described by coupled NLS equations for wave amplitudes
(envelope functions) $u(z,x)$ and $v(z,x)$ in the two cores. In the scaled
form, the NLS system is
\begin{eqnarray}
iu_{z}+\frac{1}{2}u_{xx}+|u|^{4}u+\Lambda v &=&0,  \label{u} \\
iv_{z}+\frac{1}{2}v_{xx}+|v|^{4}v+\Lambda u &=&0,  \label{v}
\end{eqnarray}%
where $z$ and $x$ are the propagation distance and transverse coordinate, $%
\Lambda $ is the constant of the inter-core tunnel coupling, and it is
assumed that the intrinsic self-focusing is represented by the quintic
terms, which may be realized in colloidal optical waveguides. As shown by
direct experiments \cite{Cid1,Cid2}, the dominance of the quintic
nonlinearity can be provided by selecting appropriate values of the size of
metallic nanoparticles and their concentration in the colloid. The analysis
clearly demonstrates that the results will not be essentially affected by a
residual cubic nonlinearity, if any.

The remaining scaling invariance of Eqs. (\ref{u}) and (\ref{v}) makes it
possible to fix $\Lambda \equiv 1$ (which implies that the propagation
distance is measured in units of the coupling length), as set below.

The emulation of SOC is provided by corrections to the linear coupling in
Eqs. (\ref{u}) and (\ref{v}) which are induced, as in Ref. \cite{NJP}, by
the skewness of the layer separating the guiding cores in the planar coupler:%
\begin{eqnarray}
iu_{z}+\frac{1}{2}u_{xx}+|u|^{4}u+v-\delta \cdot v_{x} &=&0,  \label{udelta}
\\
iv_{z}+\frac{1}{2}v_{xx}+|v|^{4}v+u+\delta \cdot u_{x} &=&0,  \label{vdelta}
\end{eqnarray}%
where real $\delta >0$ represents a shear between the cores in the skewed
coupler, see Fig. 1 in \cite{NJP}. It can be defined in terms of two
components, $u(x)$ and $v(x)$, of the full modal eigenfunction of the
dual-core waveguide: the respective overlap integral, $\int u_{\text{\textrm{%
modal}}}(x)v_{\text{\textrm{modal}}}^{\ast }(x-\Delta x)dx$ (with $\ast $
standing for the complex conjugate), attains a maximum when the inter-core
shift takes value $\Delta x=\delta $. In comparison with the basic SOC
models, the first $x$-derivatives in Eqs. (\ref{udelta}) and (\ref{vdelta})
may be considered as emulating the anomalous velocity, while the second
derivatives (the paraxial diffraction) correspond to the normal velocity
\cite{super1,super2}.

In the framework of the present model, stationary states with real
propagation constant $k$ are looked for by substituting
\begin{equation}
\left\{ u,v\right\} =e^{ikz}\left\{ U(x;\delta ),V(x;\delta )\right\}
\label{uv}
\end{equation}%
in equations (\ref{udelta}) and (\ref{vdelta}) for the wave amplitudes. The
corresponding equations for real functions $U(x;\delta )$ and $V(x;\delta )$
are%
\begin{eqnarray}
-kU+\frac{1}{2}\frac{d^{2}U}{dx^{2}}+U^{5}+V-\delta \cdot \frac{dV}{dx} &=&0,
\label{Udelta} \\
-kV+\frac{1}{2}\frac{d^{2}V}{dx^{2}}+V^{5}+U+\delta \cdot \frac{dU}{dx} &=&0.
\label{Vdelta}
\end{eqnarray}

It is relevant to identify the spectrum of the linearized version of the
present system. Looking for solutions to the linearization of Eqs. (\ref%
{Udelta}) and (\ref{Vdelta}) in the form of plane waves, \
\begin{equation}
\left\{ U,V\right\} \sim \exp \left( iqx\right)  \label{tail}
\end{equation}%
with wavenumber $q$, one derives the following dispersion relation between $%
k $ and $q^{2}$:%
\begin{equation}
k=-(1/2)q^{2}\pm \sqrt{1+\delta ^{2}q^{2}}.  \label{k}
\end{equation}%
More convenient for the subsequent analysis is an inverted form of the
relation:
\begin{equation}
q^{2}=2\left( \delta ^{2}-k\pm \sqrt{\delta ^{4}-2\delta ^{2}k+1}\right) .
\label{q^2}
\end{equation}

The spectrum determined by Eqs. (\ref{k}) and (\ref{q^2}) includes \textit{%
gaps}, i.e., intervals of propagation constant $k$ in which the plane waves
do not exist, hence they may be populated by solitons, in the framework of
the full nonlinear system. In the case of $\delta ^{2}>1$ (when the
pseudo-SOC is strong enough), we identify the semi-infinite \textit{main gap}
as
\begin{equation}
k>k_{\max }\equiv \frac{\delta ^{4}+1}{2\delta ^{2}}.  \label{semiinf}
\end{equation}%
In this gap, expressions (\ref{q^2}) are complex, hence the corresponding
values of $q$ are complex too, and solitons populating the gap will have
oscillatory tails (see Fig. \ref{fig6} below). In the case of $\delta ^{2}<1$%
, there is an additional finite \textit{annex gap}, adjacent to the main one,%
\begin{equation}
1<k<\frac{\delta ^{4}+1}{2\delta ^{2}}.  \label{annex}
\end{equation}%
In the annex gap, both branches of expression (\ref{q^2}) are real negative
ones, which makes the respective values of $q$ purely imaginary, hence the
respective solitons should have tails monotonously decaying at $%
|x|\rightarrow \infty $, see Fig. \ref{fig7} below. A given value of the
propagation constant, $k>1$, belongs to the annex gap (\ref{annex}) for
values of the pseudo-SOC strength%
\begin{equation}
\delta ^{2}<k-\sqrt{k^{2}-1}.  \label{delta^2}
\end{equation}

\begin{figure}[tbp]
\centering{\subfigure[]{\includegraphics[scale=0.40]{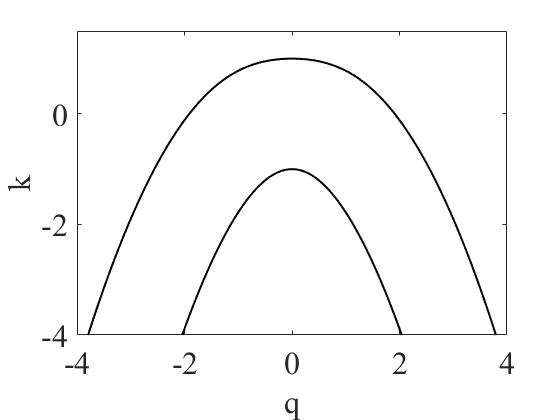}} %
\subfigure[]{\includegraphics[scale=0.40]{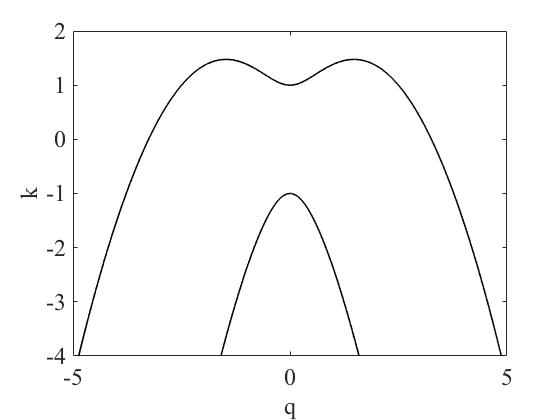}}}
\caption{The dispersion curves produced by Eq. (\protect\ref{k}) with $%
\protect\delta =0.8$ (a) and $\protect\delta =1.6$ (b).}
\label{fig0}
\end{figure}

Generic examples of the dispersion curves with $\delta ^{2}<1$ and $\delta
^{2}>1$ are displayed in Fig. \ref{fig0}. In the latter case, the largest
value $k_{\max }$ of $k$, defined by Eq. (\ref{semiinf}), is attained at $%
q=\pm \sqrt{\delta ^{2}-1/\delta ^{2}}$.

In the absence of SOC, a coupled system with the opposite (defocusing) sign
of the quintic nonlinearity, competing with cubic focusing, was considered
in work \cite{Lior}, which addressed phenomenology of spontaneous symmetry
breaking (SSB)\ of two-component solitons in the symmetric system. In that
system, symmetric solitons in the absence of SSB, as well as asymmetric ones
generated by SSB, may be stable due to the absence of the critical collapse.

The SOC-emulating terms in Eqs. (\ref{udelta}) and (\ref{vdelta}) break the
Galilean invariance of the system, therefore generation of \textquotedblleft
moving" solutions (in fact, spatial beams tilted in the $\left( x,z\right) $
plane) from their \textquotedblleft quiescent" counterparts (straight beams
in the same plane) is a nontrivial issue. For this purpose, it is relevant
to rewrite Eqs. (\ref{udelta}) and (\ref{vdelta}) in terms of $z$ and the
tilted coordinate,%
\begin{equation}
\xi \equiv x-cz,  \label{xi2}
\end{equation}%
where \textquotedblleft velocity" $c$ determines the obliquity of the beams
in the spatial domain:%
\begin{eqnarray}
iu_{z}-icu_{\xi }+\frac{1}{2}u_{\xi \xi }+|u|^{4}u+v-\delta \cdot v_{\xi }
&=&0,  \notag \\
iv_{z}-icv_{\xi }+\frac{1}{2}v_{\xi \xi }+|v|^{4}v+u+\delta \cdot u_{\xi }
&=&0.  \label{xi}
\end{eqnarray}%
Further, the "tilted" system (\ref{xi}) may be simplified by the following
transformation, which is suggested by the Galilean boost in systems which
are Galilean invariant:%
\begin{equation}
\left\{ u(\xi ,z),v(\xi ,z)\right\} \equiv \exp \left( \frac{i}{2}%
c^{2}z+ic\xi \right) \left\{ \tilde{u}(\xi ,z),\tilde{v}(\xi ,z)\right\} .
\label{tilde}
\end{equation}%
This transformation casts the equations in the following form:%
\begin{eqnarray}
i\tilde{u}_{z}+\frac{1}{2}\tilde{u}_{\xi \xi }+|\tilde{u}|^{4}\tilde{u}%
+\left( \Lambda -ic\delta \right) \tilde{v}-\delta \cdot \tilde{v}_{\xi }
&=&0,  \notag \\
i\tilde{v}_{z}+\frac{1}{2}\tilde{v}_{\xi \xi }+|\tilde{v}|^{4}\tilde{v}%
+\left( \Lambda +ic\delta \right) \tilde{u}+\delta \cdot \tilde{u}_{\xi }
&=&0.  \label{transformed}
\end{eqnarray}%
Stationary solutions to Eq. (\ref{transformed}) may be sought for as $%
\left\{ \tilde{u},\tilde{v}\right\} =e^{ikz}\left\{ \tilde{U}(\xi ),\tilde{V}%
(\xi )\right\} $, cf. Eq. (\ref{uv}), with complex functions $\tilde{U}(\xi
) $ and $\tilde{V}(\xi )$ satisfying equations
\begin{eqnarray}
-k\tilde{U}+\frac{1}{2}\frac{d^{2}\tilde{U}}{d\xi ^{2}}+\left\vert \tilde{U}%
\right\vert ^{4}\tilde{U}+\left( 1-ic\delta \right) \tilde{V}-\delta \frac{d%
\tilde{V}}{d\xi } &=&0,  \label{Utilde} \\
-k\tilde{V}+\frac{1}{2}\frac{d^{2}\tilde{V}}{d\xi ^{2}}+\left\vert \tilde{V}%
\right\vert ^{4}\tilde{V}+\left( 1+ic\delta \right) \tilde{U}+\delta \frac{d%
\tilde{U}}{d\xi } &=&0.  \label{Vtilde}
\end{eqnarray}

The underlying system of coupled NLS equations (\ref{u}) and (\ref{v})
conserves the total norm, Hamiltonian, and momentum:%
\begin{equation}
N\equiv N_{u}+N_{v}=\int_{-\infty }^{+\infty }\left[ \left\vert
u(x)\right\vert ^{2}+\left\vert v(x)\right\vert ^{2}\right] dx,  \label{norm}
\end{equation}%
\begin{gather}
H=\int_{-\infty }^{+\infty }\left[ \frac{1}{2}\left( \left\vert
u_{x}\right\vert ^{2}+\left\vert v_{x}\right\vert ^{2}\right) -\frac{1}{3}%
\left( |u|^{6}+|v|^{6}\right) \right.  \notag \\
\left. -\left( uv^{\ast }+u^{\ast }v\right) +\frac{\delta }{2}\left( u^{\ast
}v_{x}+uv_{x}^{\ast }-u_{x}^{\ast }v-u_{x}v^{\ast }\right) \right] dx,
\label{H}
\end{gather}%
\begin{equation}
P=i\int_{-\infty }^{+\infty }\left( u_{x}^{\ast }u+v_{x}^{\ast }v\right) dx.
\label{P}
\end{equation}%
Note that momentum (\ref{P}) is conserved in spite of the lack of the
Galilean invariance of the system.

\section{Asymmetric solitons in the absence of the spin-orbit coupling
(SOC), $\protect\delta =0$}

\subsection{The spontaneous-symmetry-breaking (SSB)\ point of two-component
solitons}

Before presenting results for solitons under the action of the pseudo-SOC,
it is relevant to briefly address ones in the system of Eqs. (\ref{u}) and (%
\ref{v}) in the absence of SOC. As mentioned above, in that case the issue
of basic interest is the SSB of two-component solitons in the symmetric
system \cite{Wabnitz}-\cite{Peng}.

For symmetric solitons, with $U(x)=V(x)\equiv U_{0}(x)$, Eqs. (\ref{Udelta})
and (\ref{Vdelta}) with $\delta =0$ reduce to a single equation:%
\begin{equation}
-\left( k-1\right) U_{0}+\frac{1}{2}\frac{d^{2}U_{0}}{dx^{2}}+U_{0}^{5}=0.
\label{U=V}
\end{equation}%
The exact solution to Eq. (\ref{U=V}), which represents the 1D (\textit{quasi%
}) version of the TS (Townes' soliton), exists for $k>1$:%
\begin{equation}
U_{0}(x)=\frac{\left[ 3\left( k-1\right) \right] ^{1/4}}{\sqrt{\cosh \left( 2%
\sqrt{2\left( k-1\right) }x\right) }}.  \label{exact}
\end{equation}%
In agreement with the above-mentioned fundamental property of TSs, the norm
of solution (\ref{norm}) takes a single value which does not depend on $k$:%
\begin{equation}
N_{\mathrm{TS}}=\sqrt{3/2}\pi .  \label{Norm}
\end{equation}

Only the symmetric solitons exist at $k<k_{\mathrm{cr}}$, while asymmetric
ones appear at $k>k_{\mathrm{cr}}$. The critical value $k_{\mathrm{cr}}$ can
be found exactly, by looking for a solution with a vanishingly small
asymmetric part, as%
\begin{equation}
\left( U(x),V(x)\right) =U_{0}(x)\pm \epsilon \delta U(x),  \label{+-}
\end{equation}%
where $\epsilon $ is an infinitely small amplitude, which accounts for the
onset of the SSB. Straightforward analysis, which follows that developed
earlier for the coupler with the cubic nonlinearity \cite{Wabnitz,Wabnitz2},
demonstrates that $\delta U(x)$ satisfies the following linear equation:%
\begin{equation}
\left[ \frac{1}{2}\frac{d^{2}}{dx^{2}}+\frac{15\left( k-1\right) }{\cosh
^{2}\left( 2\sqrt{2\left( k-1\right) }x\right) }\right] \delta U=\left(
k+1\right) \delta U.  \label{deltaU}
\end{equation}%
Making use of the known exact solutions for the linear Schr\"{o}dinger
equation with the P\"{o}schl-Teller potential \cite{Landau}, we find the SSB
point from a solution of Eq. (\ref{deltaU}):%
\begin{equation}
k_{\mathrm{cr}}=5/4.  \label{cr}
\end{equation}

In addition to the symmetric and asymmetric soliton solutions, Eqs. (\ref%
{Udelta}) and (\ref{Vdelta}) admit antisymmetric ones, with $U(x;\delta
)=-V(x;\delta )$. However, they are subject to strong instability, which is
driven by the fact that the coupling term in Hamiltonian (\ref{H}) is
positive for the antisymmetric solutions, on the contrary to the negative
one for the symmetric solitons \cite{Peng}. Further, the analysis similar to
that leading to Eq. (\ref{deltaU}) demonstrates that the antisymmetric
solitons do not undergo a bifurcation of antisymmetry breaking (the same
happens in the coupler with the cubic nonlinearity \cite{Peng}).

\subsection{The variational approximation (VA) for the asymmetric solitons}

At $k>k_{\mathrm{cr}}$, asymmetric solitons cannot be found in an exact
form, but it is possible to approximate them by means of the variational
method, cf. works \cite{Pare}-\cite{Skinner} for the coupler with the cubic
nonlinearity (see also a review in \cite{Peng}). To this end, we note that
Eqs. (\ref{Udelta}) and (\ref{Vdelta}) with $\delta =0$ can be derived from
the following Lagrangian:%
\begin{gather}
L=\int_{-\infty }^{+\infty }\left\{ \frac{k}{2}\left( U^{2}+V^{2}\right) +%
\frac{1}{4}\left[ \left( \frac{dU}{dx}\right) ^{2}+\left( \frac{dV}{dx}%
\right) ^{2}\right] \right.  \notag \\
\left. -\frac{1}{6}\left( U^{6}+V^{6}\right) -UV\right\} dx,  \label{L}
\end{gather}%
cf. Hamiltonian (\ref{H}) with $\delta =0$. The VA can be applied, using an
ansatz whose form is suggested by solution (\ref{exact}):%
\begin{equation}
\left( U,V\right) =\frac{A\left( \cos \theta ,\sin \theta \right) }{\sqrt{%
\cosh \left( ax\right) }},  \label{ans}
\end{equation}%
where $A$ and $a$ determine the amplitude and inverse width of the soliton,
while its asymmetry is determined by a parameter which measures the relative
difference of the norms of the two components, see Eq. (\ref{norm}):
\begin{equation}
\Theta \equiv \frac{N_{u}-N_{v}}{N_{u}+N_{v}}=\cos \left( 2\theta \right)
\label{cos}
\end{equation}%
The total norm of this ansatz, $N=\pi A^{2}/a$, does not depend on $\theta $%
. Below, the amplitude is eliminated in favor of $N$, using this relation.

The substitution of ansatz (\ref{ans}) in the Lagrangian and subsequent
integration produces the following result:%
\begin{equation}
L\left( N,a,\theta \right) =\frac{k}{2}N+\frac{Na^{2}}{32}-\frac{N^{3}a^{2}}{%
96\pi ^{2}}\left[ 5+3\cos \left( 4\theta \right) \right] -\frac{N}{2}\sin
\left( 2\theta \right) ,  \label{Leff}
\end{equation}%
The variational (Euler-Lagrange) equations, $\partial L/\partial \left(
2\theta \right) =\partial L/\partial a=\partial L/\partial N$ $=0$ yield,
after some algebra, the following results for asymmetric solitons, with $%
\cos \left( 2\theta \right) \neq 0$:
\begin{equation}
\Theta (N)=\sqrt{\left( 1/3\right) \left[ \left( N_{\mathrm{TS}}/N\right)
^{2}-1\right] },  \label{theta}
\end{equation}%
\begin{equation}
a^{2}(N)=\frac{2\pi ^{2}\sqrt{3}}{N^{2}\sqrt{1-\left( N_{\mathrm{TS}%
}/2N\right) ^{2}}},  \label{a}
\end{equation}%
\begin{equation}
k(N)=\frac{2}{\sqrt{3\left[ 1-\left( N_{\mathrm{TS}}/2N\right) ^{2}\right] }}%
.  \label{k2}
\end{equation}%
where $N_{\mathrm{TS}}$ is the 1D-TS norm (\ref{Norm}). Thus, it is expected
that the asymmetric solitons exist with norms taking values which secure the
positiveness of the expression under the square root in Eqs. (\ref{a}) and (%
\ref{k2}), and fulfillment of condition $0\leq \cos ^{2}\left( 2\theta
\right) \leq 1$, in interval
\begin{equation}
N_{\mathrm{TS}}/2<N<N_{\mathrm{TS}}.  \label{NNN}
\end{equation}%
It is easy to see that this existence interval is an exact one (without the
reference to the VA), with $\Theta \left( N=N_{\mathrm{TS}}/2\right)
=1,k\left( N=N_{\mathrm{TS}}/2\right) =\infty $, and $\Theta \left( N=N_{%
\mathrm{TS}}\right) =0$, $k\left( N=N_{\mathrm{TS}}\right) =5/4$ (see Eq. (%
\ref{cr})).

Note that the TS norm for the single 1D NLS equation with the quintic
self-focusing term (i.e., Eq. (\ref{u}) with $\Lambda =0$) is half of the
value given by Eq. (\ref{Norm}), hence development of instability of
asymmetric solitons with the norm belonging to interval (\ref{NNN}) may
result in either decay of the soliton or the onset of the collapse in one
core, see Figs. \ref{fig4} and \ref{fig5} below. In this connection, it is
relevant to mention that the VA predicts that the norm in one core
(normalized to $N_{\mathrm{TS}}/2$), which is
\begin{equation}
\frac{N_{u}}{N_{\mathrm{TS}}/2}\equiv \frac{N}{N_{\mathrm{TS}}/2}\cos
^{2}\theta =\frac{N}{N_{\mathrm{TS}}}+\sqrt{\frac{1}{3}\left( 1-\frac{N^{2}}{%
N_{\mathrm{TS}}^{2}}\right) },  \label{Nu}
\end{equation}%
pursuant to the definition of ansatz (\ref{ans}) and Eq. (\ref{theta}),
attains a maximum, $\left( 2N/N_{\mathrm{TS}}\right) _{\max }=2/\sqrt{3}$,
at $N=\left( \sqrt{3}/2\right) N_{\mathrm{TS}}$, the respective value of the
propagation constant being
\begin{equation}
k\left( N/N_{\mathrm{TS}}=1/\sqrt{3}\right) =\sqrt{2}.  \label{k=}
\end{equation}

Thus, the VA predicts that the asymmetric solitons emerge \textit{%
subcritically}, at $N=N_{\mathrm{TS}}/2$, with the largest degree of the
asymmetry, $\Theta \left( N=N_{\mathrm{TS}}/2\right) =1$, attained at the
diverging propagation constant, $k\rightarrow \infty $, as per Eq. (\ref{k2}%
). As shown in Fig. \ref{fig1}(a), with the increase of $N$ from $N_{\mathrm{%
TS}}/2$ to $N_{\mathrm{TS}}$, the asymmetry decreases, following Eq. (\ref%
{theta}), and vanishes when the norm attains the TS value, $\Theta \left(
N=N_{\mathrm{TS}}\right) =0$, the respective propagation constant, as
predicted by the variational equation (\ref{k2}), being $k_{\mathrm{cr}%
}=k\left( N=N_{\mathrm{TS}}\right) =4/3$ (to be compared to the exact value $%
5/4$, see Eq. (\ref{cr}), i.e., the relative inaccuracy of the VA is $%
\approx 6\%$). This picture, which may be identified as a \textit{fully
subcritical} SSB bifurcation (similar to the ``fully backward" one found as
an exact solution of the 1D NLS equation with the cubic self-focusing term
multiplied by a symmetric pair of delta-functions, $\delta \left( x-a\right)
+\delta \left( x+a\right) $ \cite{Thawatchai}), is drastically different
from the \textit{weakly subcritical} SSB\ bifurcation of solitons in the
coupler with the cubic self-focusing, that gives rise to a pair of
background-going branches of asymmetric unstable states, which quickly turn
forward as stable branches \cite{Skinner,Peng}. The branches of asymmetric
solitons merge with the one of symmetric states at $N=N_{\mathrm{TS}}$, and,
naturally, no solitons exist at $N>N_{\mathrm{TS}}$, as the presence of the
collapse does not allow the existence of solitons with a supercritical norm.
An essential difference from the bifurcation diagram reported in Ref. \cite%
{Thawatchai} is that, in the present system, all symmetric solitons (TSs),
having the single value of the norm, $N=N_{\mathrm{TS}}$, are represented by
the single point in Fig. \ref{fig1}(a).

The monotonously decreasing dependence $k(N)$, predicted by the VA in the
form of Eq. (\ref{k2}) (Fig. \ref{fig1}(b)), contradicts the well-known
Vakhitov-Kolokolov criterion, $dk/dN>0$, which is a necessary stability
condition for solitons \cite{VK,Berge,Fibich}, hence the entire family of
the asymmetric solitons is unstable (similar to what was found for the case
of the fully backward bifurcation in \cite{Thawatchai}).

\subsection{Numerical results for the asymmetric solitons}

\begin{figure}[tbp]
\centering{\subfigure[]{\includegraphics[scale=0.50]{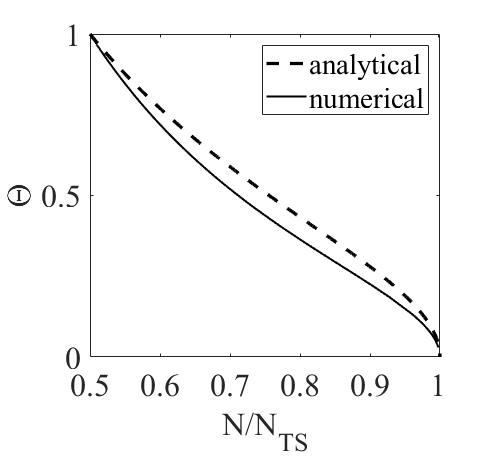}} %
\subfigure[]{\includegraphics[scale=0.50]{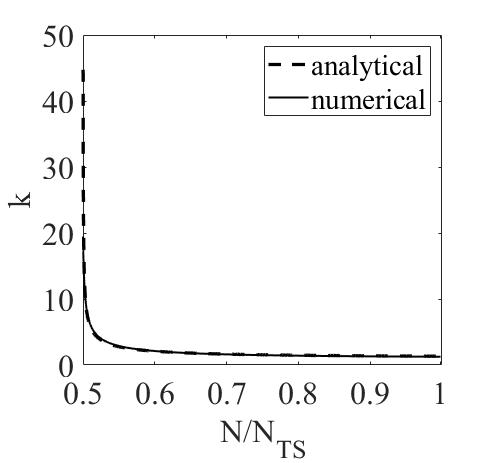}}}
\caption{{}Asymmetric solitons in the absence of SOC ($\protect\delta =0$).
(a) Asymmetry parameter $\Theta $, defined as per Eq. (\protect\ref{cos}):
the analytical prediction of the VA (variational approximation), given by
Eq. (\protect\ref{theta}), and its counterpart, produced by numerical
solution of Eqs. (\protect\ref{Udelta}) and (\protect\ref{Vdelta}), vs. the
scaled norm. In this diagram, all symmetric (\textit{quasi-Townes}')
solitons are represented by the single point, $\left( N/N_{\mathrm{TS}%
}=1,\Theta =0\right) $. (b) The propagation constant of the asymmetric
soliton vs. its total norm: the analytical VA prediction, given by Eq. (%
\protect\ref{k2}), and its numerical counterpart, vs. the scaled norm. Exact
symmetric solitons (\protect\ref{exact}) are actually represented in panel
(b) by the right edge of the frame, corresponding to $N/N_{\mathrm{TS}}=1$,
at $k>1$, which intersects the branch of the asymmetric solutions at $k=5/4$%
, according to Eq. (\protect\ref{cr}).}
\label{fig1}
\end{figure}

The predictions of the VA for the family of asymmetric solitons, \emph{viz}%
., dependences of the asymmetry parameter and propagation constant on the
total norm, are compared to their counterparts, found from the numerical
solution of Eq. (\ref{U=V}), in Fig. \ref{fig1}. The comparison corroborates
the accuracy of the approximation.

\begin{figure}[tbp]
{\subfigure[]{\includegraphics[scale=0.50]{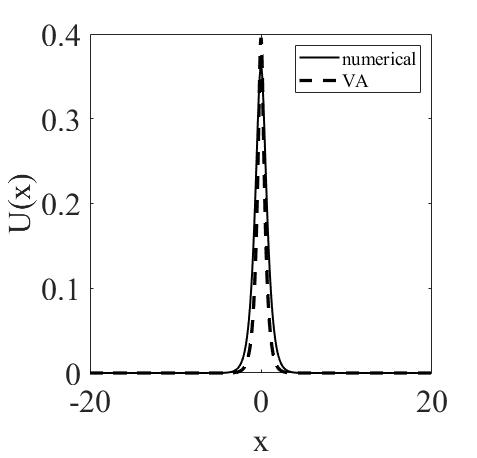}} \subfigure[]{%
\includegraphics[scale=0.50]{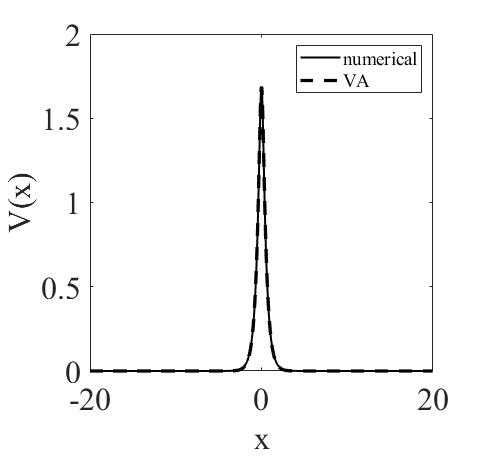}}}
\caption{{}An example of a strongly asymmetric soliton in the system without
SOC ($\protect\delta =0$), as predicted by the VA based on Eqs. (\protect\ref%
{ans}) and (\protect\ref{theta})-(\protect\ref{k2}), and found in the
numerical form. Note the difference in vertical scales between (a) and (b).
The propagation constant is $k=3$, the respective VA-predicted scaled norm
being $\left( N/N_{\mathrm{TS}}\right) _{\mathrm{VA}}\approx 0.541$, while
its numerically found counterpart is $\left( N/N_{\mathrm{TS}}\right) _{%
\mathrm{num}}\approx 0.545$.}
\label{fig2}
\end{figure}

Typical examples of strongly and moderately asymmetric solitons, as produced
by the VA and numerical solution, are displayed in Figs. \ref{fig2} and \ref%
{fig3}, and their dynamical instability is shown in Figs. \ref{fig4} and \ref%
{fig5}, respectively. The instability was identified in direct simulations,
as well as via numerical calculation of eigenvalues $\lambda $ of modes of
small perturbations, produced by the linearization of Eqs. (\ref{u}) and (%
\ref{v}) (i.e., Bogoliubov - de Gennes equations, in terms if the SOC
emulation). The perturbation modes were taken, as usual, as a combination of
terms $\sim \exp \left( -i\lambda z\right) $ and $\exp \left( i\lambda
^{\ast }z\right) $ \cite{Yang}. The \textquotedblleft normal" exponentially
growing instability is determined by $\left\vert \mathrm{Im}(\lambda
)\right\vert $ in the case when complex eigenvalues are found. On the other
hand, as mentioned above, the subexponential (but, nevertheless, tangible)
instability of TSs is accounted for by zero eigenvalues \cite{Fibich}.
\begin{figure}[tbp]
\centering{\subfigure[]{\includegraphics[scale=0.50]{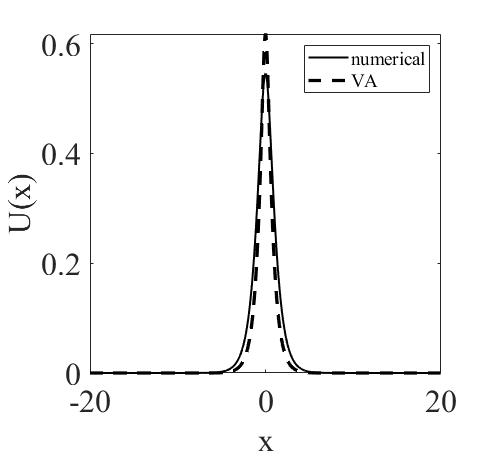}} %
\subfigure[]{\includegraphics[scale=0.50]{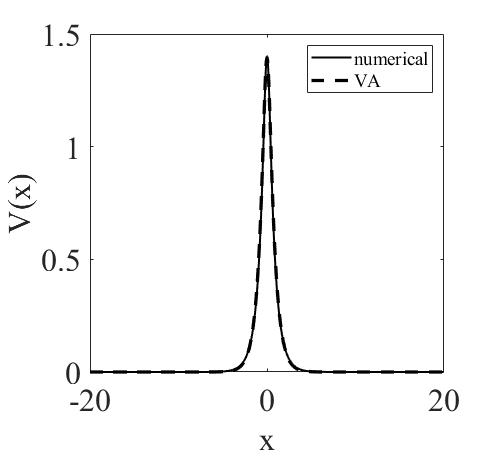}}}
\caption{{}The same as in Fig. \protect\ref{fig2}, but for a less asymmetric
soliton, corresponding to $k=1.8$. In this case, $\left( N/N_{\mathrm{TS}%
}\right) _{\mathrm{VA}}\approx 0.651$, and its numerically found counterpart
is $\left( N/N_{\mathrm{TS}}\right) _{\mathrm{num}}\approx 0.648$.}
\label{fig3}
\end{figure}
\begin{figure}[tbp]
\centering{\subfigure[]{\includegraphics[scale=0.53]{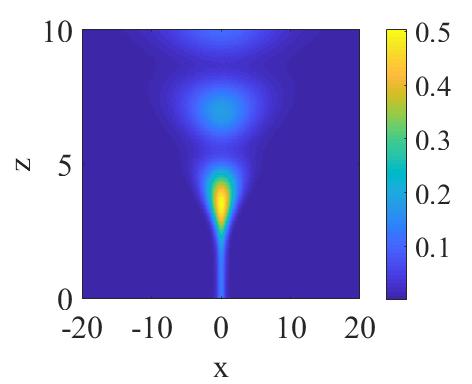}} %
\subfigure[]{\includegraphics[scale=0.53]{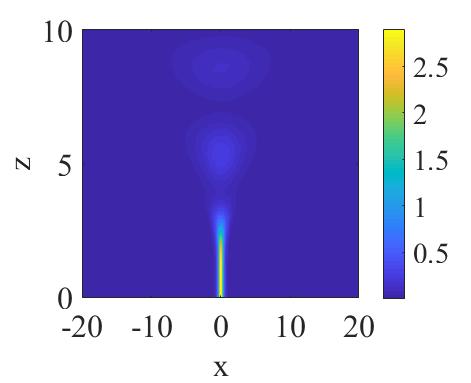}} \subfigure[]{%
\includegraphics[scale=0.53]{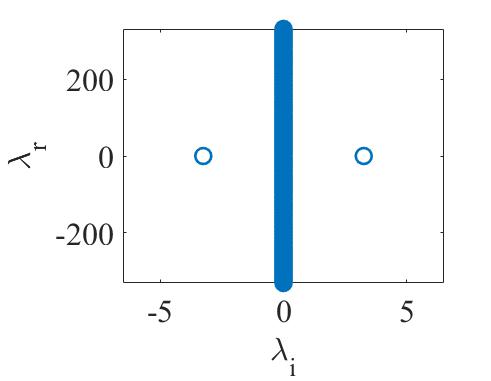}}}
\caption{{}The instability of the asymmetric soliton from Fig. \protect\ref%
{fig2}. Panels (a) and (b) display, by means of contour plots, the decay of
components $\left\vert u\left( x,z\right) \right\vert ^{2}$ and $\left\vert
v\left( x,z\right) \right\vert ^{2}$, respectively. (c) The spectrum of
eigenvalues of small perturbations added to the stationary soliton, as
produced by the numerical solution of the linearized equations for small
perturbations.}
\label{fig4}
\end{figure}
\begin{figure}[tbp]
\centering{\subfigure[]{\includegraphics[scale=0.60]{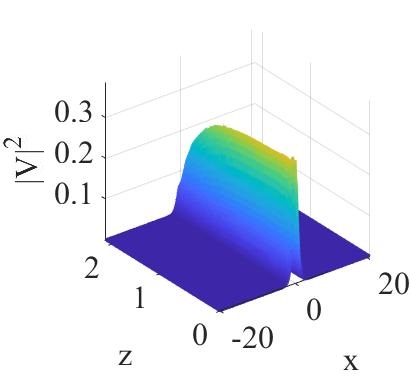}} %
\subfigure[]{\includegraphics[scale=0.60]{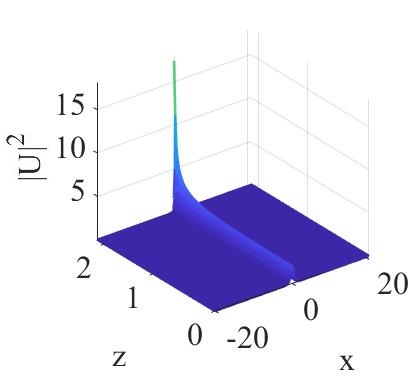}}} \subfigure[]{%
\includegraphics[scale=0.3]{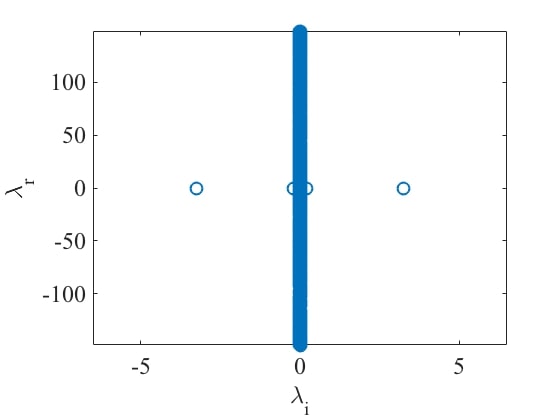}}
\caption{{}The collapse instability of the soliton from Fig. \protect\ref%
{fig3}. The meaning of the panels is the same as in Fig. \protect\ref{fig4},
with the difference that the evolution of the two components is displayed,
in panels (a) and (b), by means of the three-dimensional plots (note the
difference in vertical scales in (a) and (b)).}
\label{fig5}
\end{figure}

In Fig. \ref{fig4} it is observed that the instability of the strongly
asymmetric soliton with a relatively small norm leads to decay, while the
less asymmetric one with a higher norm suffers the collapse in Fig. \ref%
{fig5}. This difference is explained by the fact that, in the former case
(with $k=3$), Eq. (\ref{Nu}) yields the norm in the dominant core which only
slightly exceeds the single-core critical value: $N_{u}/\left( N_{\mathrm{TS}%
}/2\right) \approx 1.027$. On the other, in the latter case (with $k=1.8$,
which is closer to the \textquotedblleft most unstable" value (\ref{k=})),
the same ratio is $\approx 1.090$, making it possible to initiate the
collapse, as observed in Fig. \ref{fig5}.

In the next section, we demonstrate that the inclusion of the SOC terms in
Eqs. (\ref{u}) and (\ref{v}) may stabilize the symmetric TSs which are
subject to the above-mentioned subexponential instability, and thus create
the missing ground state in the system. On the other hand, it is not
expected that the SOC is able to suppress the strong instability of the
asymmetric solitons, which is accounted for by nonzero imaginary parts of
eigenvalues $\lambda $, see Figs. \ref{fig4}(c) and \ref{fig5}(c).
Therefore, in the subsequent analysis we focus on the stabilization of
solitons which are symmetric TSs in the absence of SOC.

\section{Stabilization of skew-symmetric solitons by the emulated SOC}

\subsection{The definition of the skew symmetry}

It is easy to see that stationary equations (\ref{Udelta}) and (\ref{Vdelta}%
) including the SOC-emulating terms with $\delta >0$ are compatible with the
\textit{skew-symmetry} constraint,%
\begin{equation}
U(x;\delta )=V(-x;\delta ),  \label{skew}
\end{equation}%
which replaces the symmetry condition occurring in the absence of SOC. This
constraint is somewhat similar to skew and cross symmetries considered in
models based on lattice potentials \cite{skewsymm,cross}. In other words,
constraint (\ref{skew}) means that $U\left( x,\delta \right) +V\left(
x,\delta \right) $ and $U\left( x,\delta \right) -V\left( x,\delta \right) $
are, respectively, even and odd functions of $x$.

Unlike the symmetric solitons available in the analytical form (\ref{exact}%
), no exact solutions can be found in the presence of $\delta >0$, However,
treating the SOC strength $\delta $ as a small parameter, it is possible to
look for solutions of Eqs. (\ref{Udelta}) and (\ref{Vdelta}) perturbatively,%
\begin{equation}
\left\{ U(x),V(x)\right\} =\left\{ U_{0}(x),V_{0}\right\} +\delta \left\{
U_{1}(x),V_{1}(x)\right\} ,  \label{delta}
\end{equation}%
where $\left\{ U_{0}(x),V_{0}(x)\right\} $ is the solution (not necessarily
the symmetric one) for $\delta =0$. The substitution of ansatz (\ref{delta})
in Eqs. (\ref{Udelta}) and (\ref{Vdelta}) and linearization with respect to
small perturbations leads to the inhomogeneous equations:%
\begin{eqnarray}
\left( \frac{1}{2}\frac{d^{2}}{dx^{2}}-k+5U_{0}(x)\right) U_{1}+V_{1} &=&%
\frac{dV_{0}}{dx},  \notag \\
\left( \frac{1}{2}\frac{d^{2}}{dx^{2}}-k+5V_{0}(x)\right) V_{1}+U_{1} &=&-%
\frac{dU_{0}}{dx}.  \label{linear}
\end{eqnarray}%
Applying $d/dx$ to Eqs. (\ref{Udelta}) and (\ref{Vdelta}), and comparing the
results to Eqs. (\ref{linear}), it is easy to find an \emph{exact solution}
of the latter equations, which contains an arbitrary parameter representing
an infinitesimal shift of the soliton's center of mass. Keeping the shift
equal to zero, one obtains the following exact solution to Eqs. (\ref{linear}%
):
\begin{eqnarray}
U_{1}(x) &=&-\frac{N_{v}\delta }{N_{u}+N_{v}}\frac{dU_{0}}{dx},  \notag \\
V_{1}(x) &=&\frac{N_{u}\delta }{N_{u}+N_{v}}\frac{dV_{0}}{dx},  \label{11}
\end{eqnarray}%
where $N_{u,v}$ are norms of the two components at $\delta =0$, see Eq. (\ref%
{norm}). Actually, as said above, the relevant case (when the solitons may
be stabilized by SOC) is the one with the symmetric soliton at $\delta =0$,
In this case, solution (\ref{11}) simplifies to%
\begin{equation}
U_{1}(x)=-V_{1}(x)=-\frac{\delta }{2}\frac{dU_{0}}{dx},  \label{delta-symm}
\end{equation}%
with $U_{0}(x)$ taken as per Eq. (\ref{exact}). This analytical
approximation is compared to the numerical results (for $\delta =0.35$,
which is not a very small value) below in Fig. \ref{fig7}(c,d).

\subsection{Skew-symmetric solitons and their stability}

\begin{figure}[tbp]
\centering{\subfigure[]{\includegraphics[scale=0.40]{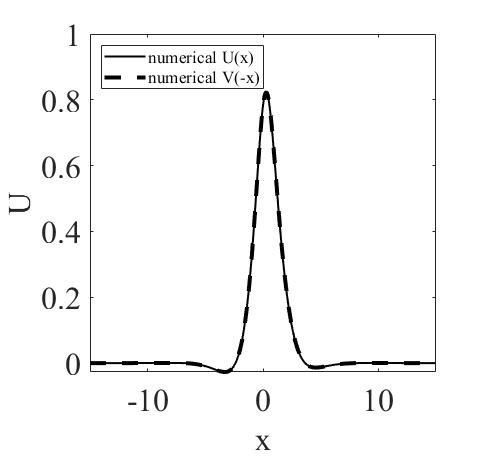}} %
\subfigure[]{\includegraphics[scale=0.40]{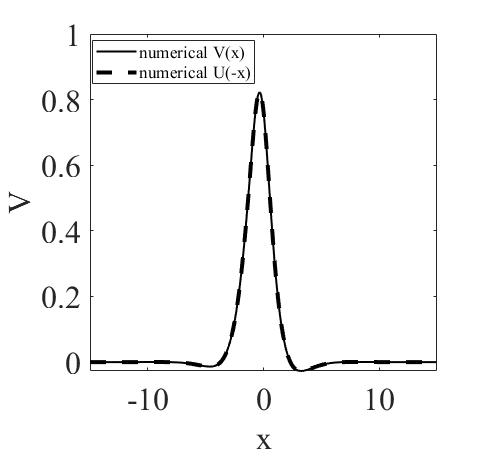}}}
\caption{{}The profile of a stable skew-symmetric soliton, corresponding to
points A ($\protect\delta =1$) in Figs. \protect\ref{fig8}(a,b). The soliton
belongs to the main spectral gap. Its propagation constant and total norm
are $k=1.2$ and $N=2.29$. The juxtaposition of profiles $U(\pm x)$ and $%
V\left( \mp x\right) $ in panels (a) and (b) corroborates that the soliton
obeys the skew-symmetry constraint (\protect\ref{skew}). }
\label{fig6}
\end{figure}

\begin{figure}[tbp]
\centering{\subfigure[]{\includegraphics[scale=0.50]{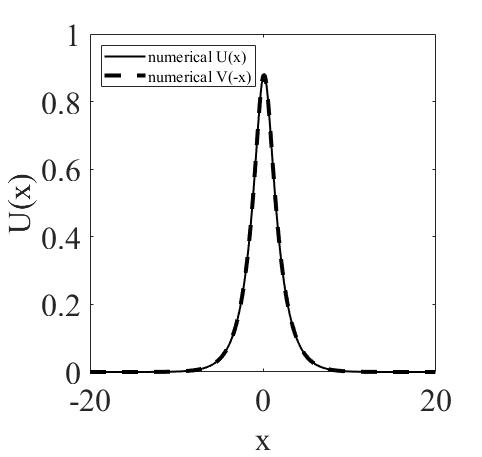}} %
\subfigure[]{\includegraphics[scale=0.50]{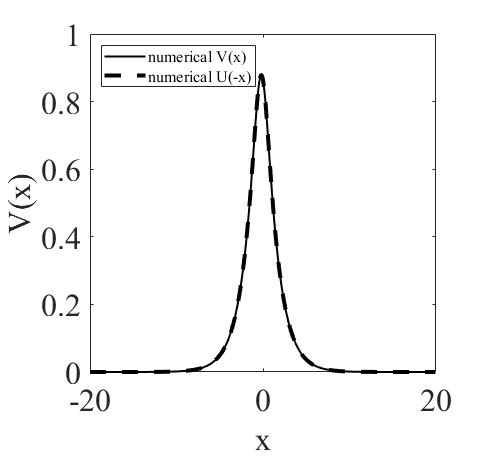}} \subfigure[]{%
\includegraphics[scale=0.50]{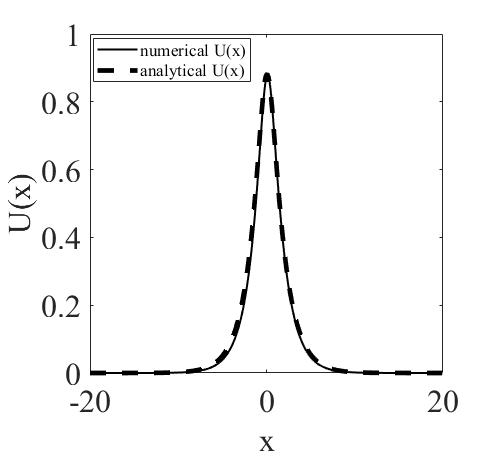}} \subfigure[]{%
\includegraphics[scale=0.50]{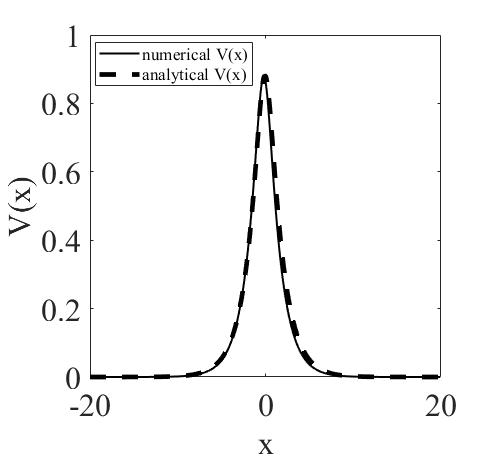}}}
\caption{{}The profile of a stable skew-symmetric soliton corresponding to
points B ($\protect\delta =0.35$) in Figs. \protect\ref{fig8}(a,b). The
soliton belongs to the annex gap. Its propagation constant and total norm
are $k=1.2$ and $N=3.61$. (a,b) The juxtaposition of profiles $U(\pm x)$ and
$V\left( \mp x\right) $ in panels (a) and (b) corroborates that the soliton
obeys the skew-symmetry constraint (\protect\ref{skew}). (c,d) Comparison
between the analytical apporixmation given by Eqs. (\protect\ref{delta}), (%
\protect\ref{delta-symm}) and the numerical solution of Eqs. (\protect\ref%
{Vdelta}) and (\protect\ref{Udelta}). To estimate the relative size of the
analytical correction given by Eq. (\protect\ref{delta-symm}), we note that,
at the point where the local intensity of solution (\protect\ref{exact}), $%
U_{0}^{2}(x)$, is half of its maximum value, Eq. (\protect\ref{delta-symm})
with $\protect\delta =0.35$ and $k=1.2$ yields $\left\vert
U_{1}/U_{0}\right\vert \approx 0.1$.}
\label{fig7}
\end{figure}

Localized solutions of coupled equations (\ref{Udelta}) and (\ref{Vdelta})
obeying condition (\ref{skew}) were obtained in a numerical form. Generic
examples of stable solitons belonging to the main and annex bandgaps,
defined as per Eqs. (\ref{semiinf}) and (\ref{annex}), are presented in
Figs. \ref{fig6} and \ref{fig7}, respectively. The former solution features
zero crossings, in accordance with the fact that, as mentioned above,
wavenumber $q$ of its tail, given by Eqs. (\ref{tail}) and (\ref{q^2}), is
complex, $q=q_{r}+iq_{i}$, hence the solution decays non-monotonously at $%
|x|\rightarrow \infty $: $\left\{ U,V\right\} \sim \exp \left( -\left\vert
q_{i}\right\vert |x|\right) \cos \left( q_{r}x\right) $. The first
zero-crossing is predicted by this estimate at
\begin{equation}
|x_{0}|=\pi /\left( 2\left\vert q_{r}\right\vert \right) .  \label{zero}
\end{equation}%
In the case shown in Fig. \ref{fig6}, with $k=1.2$, Eq. (\ref{q^2}) yields $%
q_{r}\approx 0.81$ and $q_{i}\approx 0.93$. The first zero-crossing is
observed in Fig. \ref{fig6} at $\left\vert x_{0}\right\vert \approx 2.34$,
while Eq. (\ref{zero}) gives $\left\vert x_{0}\right\vert \approx 2.3$, in
agreement with the numerical findings.{\Large \ }On the other hand, Eq. (\ref%
{q^2}) yields purely imaginary $q$ in the annex bandgap (\ref{annex}), hence
the respective soliton's shape features monotonous decay of the tails at $%
|x|\rightarrow \infty $ and, accordingly, no zero crossings are observed in
Fig. \ref{fig7}.

\begin{figure}[tbp]
\centering{\subfigure[]{\includegraphics[scale=0.4]{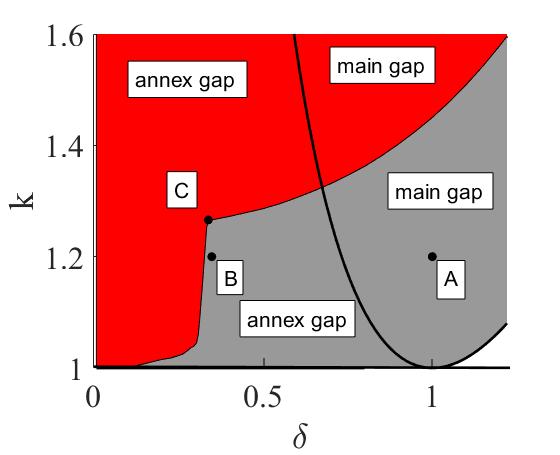}} %
\subfigure[]{\includegraphics[scale=0.4]{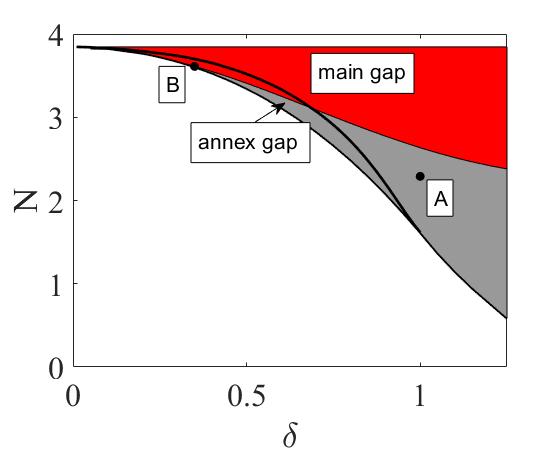}} \subfigure[]{%
\includegraphics[scale=0.4]{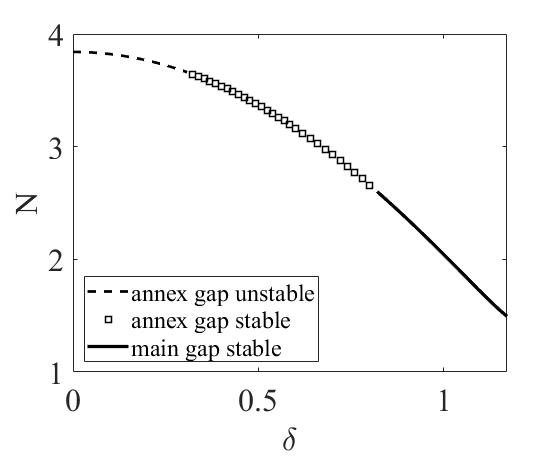}} \subfigure[]{%
\includegraphics[scale=0.40]{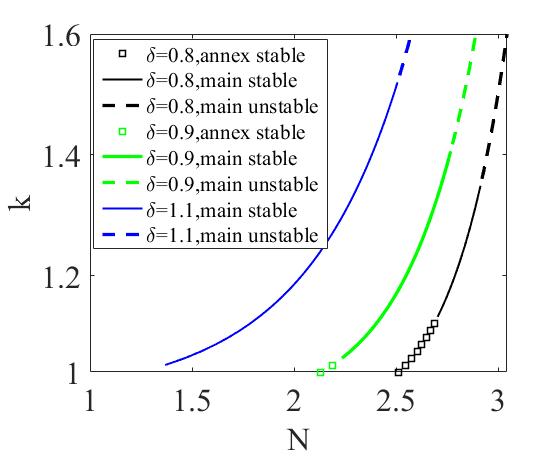}}}
\caption{{}(a,b): Existence and stability areas for skew-symmetric solitons
in the plane of the pseudo-SOC strength ($\protect\delta $) and propagation
constant ($k$) or total norm ($N$). The parabola-shaped black continuous
curve in (a), $k=\left( \protect\delta ^{4}+1\right) /\left( 2\protect\delta %
^{2}\right) $, see Eqs. (\protect\ref{semiinf}) and (\protect\ref{annex}),
is the boundary between the main and annex spectral gaps at $\protect\delta %
<1$, and a border of the main bandgap at $\protect\delta >1$, where the
annex gap does not exist. The boundary between the annex and main gaps is
shown in (b) too. Points A and B in (a) and (b) correspond to stable
solitons displayed in Figs. \protect\ref{fig6} and \protect\ref{fig7},
respectively. Symbol C denotes the corner point of the stabulity boundary,
with coordinate $k\approx k_{\mathrm{cr}}=5/4$, see details in the text. The
skew-symmetric solitons are unstable, stable, and nonexistent (or not found)
in the red, gray, and white areas, respectively. (c) Dependence $N(\protect%
\delta )$ for the skew-symmetric solitons at a fixed value of the
propagation constant, $k=1.1$. Pursuant to Eq. (\protect\ref{delta^2}),
segment $\protect\delta <0.801$ belongs to the annex bandgap. (d) Dependence
$k(N)$ at fixed values of the pseudo-SOC strength: $\protect\delta =0.8$; $%
0.9$; $1.1$.}
\label{fig8}
\end{figure}

Results for the existence and stability of the skew-symmetric solitons are
summarized in Fig. \ref{fig8}. The stability was identified by means of
numerical calculation of eigenvalues for modes of small perturbations, and
verified by direct simulations. In particular, the bottom existence boundary
in panel \ref{fig8}(a) is determined by Eqs. (\ref{annex}) and (\ref{semiinf}%
). Further, the coordinate $k\approx 1.265$ of corner point $C$ on the
boundary between the stability and instability areas in the same panel may
be explained by the above analytical result, according to which the
symmetry-breaking instability of the solitons in the absence of SOC sets it
at point (\ref{cr}), i.e., $k=1.25$.

The key point, which explains the stabilization of the quasi-TS by SOC as a
result of pushing the soliton's norm below its critical value for the TSs in
the absence of SOC (see Eq. (\ref{Norm})), is illustrated by Figs. \ref{fig8}%
(c,d), which show that, for a fixed propagation constant, the norm of the
skew-symmetric solitons indeed decreases with the increase of the SOC
strength, $\delta $, leading to the stabilization of the solitons. In this
connection, it is relevant to mention that, as long as $k$ belongs to the
annex gap (\ref{annex}), in the limit of $k\rightarrow 1$ the stationary
equations may be asymptotically reduced to Eqs. (\ref{Udelta}), (\ref{Vdelta}%
) with $\delta =0$ and the effective diffractive coefficient, taken from the
expansion of the top branch of the dispersion relation (\ref{k}) at $%
k\rightarrow 1$:
\begin{equation*}
D\equiv -\frac{d^{2}k}{dq^{2}}|_{q=0}=1-\delta ^{2}.
\end{equation*}
Then, the total norm of the soliton can be obtained from the TS value (\ref%
{Norm}) by dint of straightforward renormalization:%
\begin{equation}
N_{\mathrm{asympt}}\left( k\rightarrow 1;\delta \right) =\sqrt{1-\delta ^{2}}%
N_{\mathrm{TS}}=\sqrt{1-\delta ^{2}}\sqrt{3/2}\pi .  \label{k->0}
\end{equation}%
In particular, for $\delta =0.8$, Eq. (\ref{k->0}) yields $N\left(
k\rightarrow 1;\delta =0.8\right) \approx \allowbreak 2.31$, which is very close
to the corresponding numerical value in Fig. \ref{fig8}(d), $N_{\mathrm{numer%
}}\left( k\rightarrow 1;\delta =0.8\right) \approx 2.32$.

Equations (\ref{Udelta}) and (\ref{Vdelta}) are also compatible with the
constraint of the \textit{skew antisymmetry},
\begin{equation}
U(x;\delta )=-V(-x;\delta ),  \label{antiskew}
\end{equation}%
cf. Eq. (\ref{skew}) (in other words, the corresponding combinations $%
U(x)+V(x)$ and $U(x)-V(x)$ are, respectively, odd and even functions of $x$%
). Because solutions obeying Eq. (\ref{antiskew}) extend the above-mentioned
strongly unstable antisymmetric solitons existing at $\delta =0$, it is
expected that the skew-antisymmetric solitons inherit the strong
instability, therefore they are not considered in this section.
Nevertheless, they are addressed in the next section dealing with the
simplified diffractionless system, where skew-symmetric solutions do not
exist.

\subsection{Tilted skew-symmetric solitons}

\begin{figure}[tbp]
\centering{\subfigure[]{\includegraphics[scale=0.32]{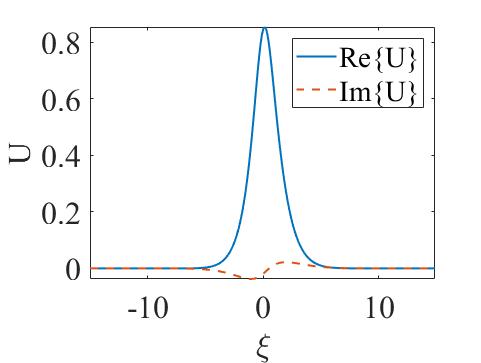}} %
\subfigure[]{\includegraphics[scale=0.32]{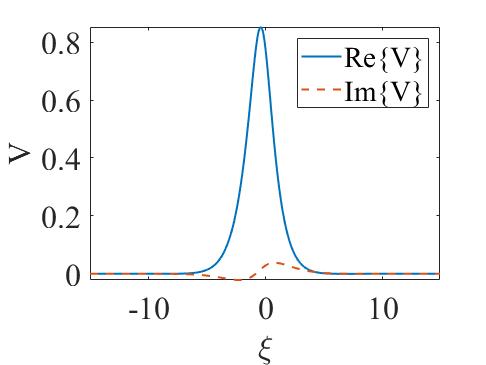}} \subfigure[]{%
\includegraphics[scale=0.37]{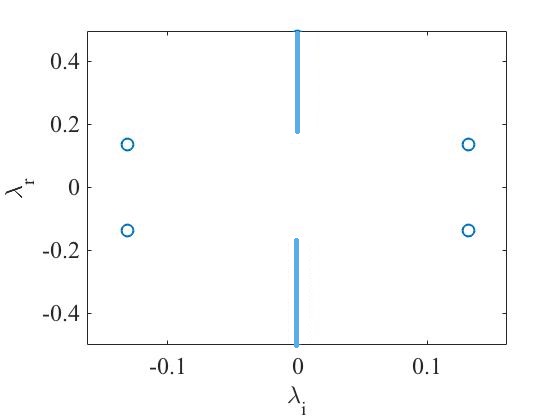}} \subfigure[]{%
\includegraphics[scale=0.30]{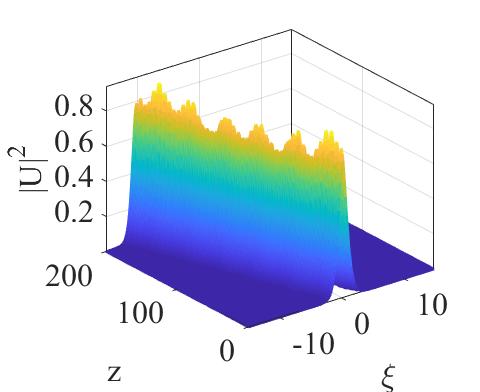}} \subfigure[]{%
\includegraphics[scale=0.30]{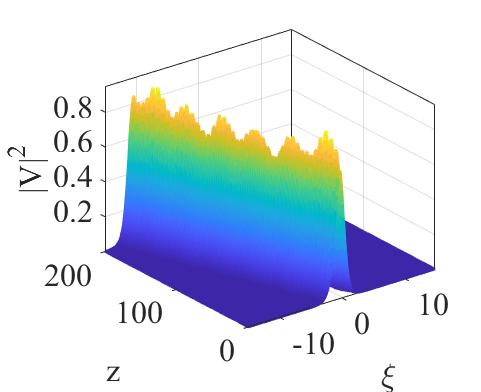}} }
\caption{{}(a,b): Complex profiles of two components of a soliton with tilt
(\textquotedblleft velocity") $c$, as found in the numerical form from Eqs. (%
\protect\ref{Utilde}) and (\protect\ref{Vtilde}). Its propagation constant,
pseudo-SOC strength and \textquotedblleft velocity" are $k=1.2$, $\protect%
\delta =0.8$ and $c=0.05$, respectively (c) The spectrum of stability
eigenvalues for pertubation modes around the soliton. The vertical stripe
with a gap represents the continuous spectrum of real eigenvalues. The
quartet of compex eigenvalues accounts for the weak instability which
transforms the stationary soliton into a breather. (d,e) The weakly unstable
evolution of the soliton shown in (a,b). }
\label{fig9}
\end{figure}

\textquotedblleft Moving" (tilted) solitons have been found as numerical
solutions of Eqs. (\ref{Utilde}) and (\ref{Vtilde}), see Figs. \ref{fig9}%
(a,b). Unlike the \textquotedblleft quiescent" (straight) solitons, the
calculation of eigenvalues in the framework of the respective linearized
equations demonstrates that all tilted solitons are unstable, see Fig. \ref%
{fig9}(c). Nevertheless, for sufficiently small values of the
\textquotedblleft velocity" (tilt) $c$, the instability is weak, the
respective eigenvalues being complex in Fig. \ref{fig9}(c), rather than
purely imaginary, cf. Figs. \ref{fig4}(c) and \ref{fig5}(c). As shown in
Figs. \ref{fig9}(d,e), in direct simulations the weak instability does not
destroy the solitons, but rather transforms them into breathers which
feature robust intrinsic vibrations, with period $Z_{\mathrm{breather}}^{%
\mathrm{(numer)}}\approx 40$ in this example. Note that the real part of the
complex eigenvalues in Fig. \ref{fig9}(e), $\lambda _{r}\simeq \pm 0.136$,
correctly predicts the period of the intrinsic vibrations in panels \ref%
{fig9}(a,b), as $Z_{\mathrm{breather}}^{\mathrm{(analyt)}}=2\pi /\left\vert
\lambda _{r}\right\vert \approx 46$. It is also relevant to compare the
vibration period with a characteristic diffraction length, $Z_{\mathrm{diffr}%
}\simeq 8$, determined by the input soliton profile, the result being $Z_{%
\mathrm{breather}}\simeq 5Z_{\mathrm{diffr}}$ (i.e., the vibrations may be
considered as a long-period dynamical feature). Figures \ref{fig9}(a,b)
additionally demonstrate secondary small-amplitude short-period oscillations
on top of the primary vibrations. This effect may be construed as the fifth
harmonic of the fundamental frequency $\lambda _{r}$, generated by the
quintic nonlinearity of the system.

At larger values of tilt $c$, solitons are subject to stronger instability,
which quickly destroys them in direct simulations (not shown here in detail).

\section{Gap solitons in the diffractionless limit}

\subsection{The simplified system}

In the case when the effective diffraction induced by the SOC terms in Eqs. (%
\ref{udelta}) and (\ref{vdelta}) is much stronger than the direct paraxial
diffraction (in terms of the SOC emulation, this condition means that the
anomalous velocity dominates over the normal one), the underlying system may
be reduced to a simpler form, in which the second derivatives are dropped,
cf. Refs. \cite{GZ} and \cite{Fukuoka}:%
\begin{eqnarray}
iu_{z}-v_{x}+|u|^{4}u+v &=&0,  \label{ugap} \\
iv_{z}+u_{x}+|v|^{4}v+u &=&0.  \label{vgap}
\end{eqnarray}%
In these equations, $\delta $ is eliminated by obvious rescaling, once the
second derivatives are absent. Accordingly, stationary real equations (\ref%
{Udelta}) and (\ref{Vdelta}) are replaced by%
\begin{eqnarray}
\frac{dV}{dx} &=&-kU+U^{5}+V,  \label{Ulimit} \\
\frac{dU}{dx} &=&kV-V^{5}-U.  \label{Vlimit}
\end{eqnarray}%
This approximation applies to localized solutions of the underlying system
with size $L\sim \delta \gg 1$ (if the coupling constant $\Lambda $ is not
set to be $\Lambda =1$ by rescaling, see Eqs. (\ref{u}) and (\ref{v}), the
conditions for the applicability of the approximation are $L\sim \delta
/\Lambda $, with $\delta \gg \sqrt{\Lambda }$).

The dispersion relation of system (\ref{ugap}), (\ref{vgap}) is also
simplified in comparison with its counterpart (\ref{k}):
\begin{equation}
k^{2}=1+q^{2}.  \label{q}
\end{equation}%
It gives rise to a finite bandgap $k^{2}<1$, in which \textit{gap solitons}
may be sought for \cite{gapsol1,gapsol2}.\ In the framework of the full
system of Eqs. (\ref{udelta}) and (\ref{vdelta}), this gap may overlap with
\textquotedblleft remote" branches of the full dispersion relation, but this
circumstance will only give rise to to decay of the gap solitons at an
exponentially small rate.

For tilted solitons, the replacement of $x$ by coordinate (\ref{xi2})
transforms Eqs. (\ref{ugap}), (\ref{vgap}) into%
\begin{eqnarray}
iu_{z}-icu_{\xi }-v_{\xi }+|u|^{4}u+v &=&0,  \notag \\
iv_{z}-icv_{\xi }+u_{\xi }+|v|^{4}v+u &=&0.  \label{tilted-limit}
\end{eqnarray}%
The bandgap generated by Eqs. (\ref{tilted-limit}) is $k^{2}<1-c^{2}$.
Obviously, gap solitons cannot exists for $c^{2}>1$. Detailed consideration
of solutions to system (\ref{tilted-limit})\ is left beyond the scope of the
present work.

\subsection{Analytical relations}

As well as solutions to the system of full equations (\ref{Udelta}) and (\ref%
{Vdelta}), Eqs. (\ref{Ulimit}) and (\ref{Vlimit}) are compatible with
skew-symmetry and antisymmetry constraints, defined by Eq. (\ref{skew}) and (%
\ref{antiskew}). Further, dividing Eq. (\ref{Ulimit}) by Eq. (\ref{Vlimit})
leads to an equations relating fields $U$ and $V$:%
\begin{equation}
\frac{dU}{dV}=-\frac{U-kV+V^{5}}{V-kU+U^{5}}.  \label{UV}
\end{equation}%
Obvious integration of Eq. (\ref{UV}) yields the following result:%
\begin{equation}
h\equiv UV-\frac{k}{2}\left( U^{2}+V^{2}\right) +\frac{1}{6}\left(
U^{6}+V^{6}\right) =0  \label{integral}
\end{equation}%
(in the general case, Eq. (\ref{integral}) has an arbitrary constant on the
right-hand side; the constant is zero for solitons, which have $U(x)$ and $%
V(x)$ vanishing at $|x|\rightarrow \infty $). In fact, expression (\ref%
{integral}) is the formal Hamiltonian of Eqs. (\ref{Ulimit}) and (\ref%
{Vlimit}), the respective canonical representation of these equations being $%
dV/dx=\partial h/\partial U,~dU/dx=-\partial h/\partial V$.

It immediately follows from Eq. (\ref{integral}) that solitons subject to
the skew-symmetry constraint (\ref{skew}) \emph{cannot exist}, because, at $%
x=0$, where Eq. (\ref{skew}) yields $U(x=0)=V(x=0)$, Eq. (\ref{integral})
amounts to%
\begin{equation}
\left( 1-k\right) U^{2}(x=0)+\frac{1}{3}U^{6}(x=0)=0,  \label{no}
\end{equation}%
which, obviously, admits solely $U(x=0)=V(x=0)=0$ inside the bandgap, at $%
k^{2}<1$, and only the trivial solution, $U=V\equiv 0$, may have both
components vanishing at $x=0$. On the other hand, the skew-antisymmetric
constraint (\ref{antiskew}) (which implies $V(x=0)=-U(x=0)$) admits the
existence of solitons, yielding the following exact value of the fields at $%
x=0$:
\begin{equation}
U(x=0)=-V(x=0)=\left[ 3\left( 1+k\right) \right] ^{1/4}.  \label{U0}
\end{equation}%
Note that the values given by Eq. (\ref{U0}) do not correspond to maxima of $%
U$ or $|V|$ (i.e., the maxima are not located at $x=0$). The maximum of $U$
(but not of $|V|$) corresponds to $dU/dx=0$, hence Eq. (\ref{Vlimit})
yields, in this case,%
\begin{equation}
-kV+V^{5}+U=0.  \label{max}
\end{equation}%
The corresponding values, $U_{\max }$ and $V$ (which does not represent a
maximum if $|V|$) should be found from the system of Eqs. (\ref{integral})
and (\ref{max}). In the general case, it is not possible to solve this
system analytically. However, a solution is available in the case of $k=0$,
i.e., at the midpoint of the finite gap:%
\begin{equation}
U_{\max }(k=0)=5^{5/24}\approx 1.398,~V(k=0)=-5^{1/24}\approx -1.069.
\label{k=0}
\end{equation}%
Accordingly, at the point of the maximum of $\left\vert V(x)\right\vert $
the values are $V_{\max }(k=0)=-5^{5/24}\approx
-1.398,~U(k=0)=5^{1/24}\approx 1.069$. For comparison, at the same value of
the propagation constant, $k=0$, Eq. (\ref{U0}) yields a value which is
smaller than $U_{\max }(k=0)$:
\begin{equation}
U(x=0,k=0)=-V(x=0,k=0)=3^{1/4}\approx 1.316.  \label{x=0}
\end{equation}

Note that Eq. (\ref{integral}) demonstrates that the solution cannot have
zeros in one component ($V=0$ or $U=0$) at $k<0$. However, zeros may exist
at $k>0$. Indeed, setting $V=0$ in Eq. (\ref{integral}), one finds that, at
this point, the value of $U$ is%
\begin{equation}
U(V=0)=\left( 3k\right) ^{1/4}  \label{V=0}
\end{equation}%
(accordingly, $V(U=0)=-\left( 3k\right) ^{1/4}$), which is relevant at $k>0$.

Further, the linearized version of Eqs. (\ref{Ulimit}) and (\ref{Vlimit})
yields the following expressions for asymptotic tails of gap solitons:%
\begin{equation}
U(x)=U_{0}\exp \left( -\sqrt{1-k^{2}}|x|\right) \left\{
\begin{array}{c}
1,\mathrm{at}~x\rightarrow +\infty , \\
-k^{-1}\left( 1-\sqrt{1-k^{2}}\right) ,~\mathrm{at}~x\rightarrow -\infty ,%
\end{array}%
\right. ,
\end{equation}%
\begin{equation}
V(x)=U_{0}\exp \left( -\sqrt{1-k^{2}}|x|\right) \left\{
\begin{array}{c}
k^{-1}\left( 1-\sqrt{1-k^{2}}\right) ,~\mathrm{at}~x\rightarrow +\infty , \\
-1,~\mathrm{at}~x\rightarrow -\infty ,%
\end{array}%
\right. ,  \label{tails}
\end{equation}%
with the exponential-decay rate, $\sqrt{1-k^{2}}$, determined by Eq. (\ref{q}%
), while constant $U_{0}$ is indefinite, in terms of the asymptotic
approximation. These expressions comply with the skew-antisymmetry condition
(\ref{antiskew}), and they have a singularity at $k\rightarrow 0$. In the
latter limit, Eqs. (\ref{tails}) amount to%
\begin{eqnarray}
V_{0} &\approx &\left( k/2\right) U_{0},~\mathrm{at}~x>0,  \label{x>0} \\
U_{0} &\approx &\left( k/2\right) V_{0},~\mathrm{at}~x<0,  \label{x<0}
\end{eqnarray}%
i.e., at $k=0$ the asymptotic tail (\ref{tails}) of either field vanishes at
positive or negative $x$. In this case, the correct asymptotic form of the
solution, replacing Eqs. (\ref{tails}), is%
\begin{equation}
U(x)=\left\{
\begin{array}{c}
U_{0}e^{-x},\mathrm{at}~x\rightarrow +\infty , \\
\left( U_{0}^{5}/6\right) e^{5x},~\mathrm{at}~x\rightarrow -\infty ,%
\end{array}%
\right.
\end{equation}%
\begin{equation}
V(x)=\left\{
\begin{array}{c}
-\left( U_{0}^{5}/6\right) e^{-5x},\mathrm{at}~x\rightarrow +\infty , \\
-U_{0}e^{x},~\mathrm{at}~x\rightarrow -\infty ,%
\end{array}%
\right.  \label{k=0tails}
\end{equation}%
where $U_{0}$ remains an indefinite constant.

\subsection{Soliton solutions}

There is a possibility to construct approximate analytical solutions for the
skew-antisymmetric gap solitons near the bottom edge of the bandgap, i.e.,
for
\begin{equation}
0<1+k\equiv \varepsilon \ll 1.  \label{<<}
\end{equation}%
In this case, it is convenient to split each component in spatially even and
odd parts:%
\begin{equation}
\left\{ U(x),V(x)\right\} =\left\{ U_{\mathrm{even}}(x),V_{\mathrm{even}%
}(x)\right\} +\left\{ U_{\mathrm{odd}}(x),V_{\mathrm{odd}}(x)\right\} .
\label{eo}
\end{equation}%
Then, the consideration of Eqs. (\ref{Ulimit}) and (\ref{Vlimit}) suggests
that, in the case of small $\varepsilon $, the solitons have large width $W$
and small amplitudes of the even and odd parts, $A_{\mathrm{even,odd}}$,
estimated as%
\begin{equation}
W\sim \varepsilon ^{-1/2},~A_{\mathrm{even}}\sim \varepsilon ^{1/4},~A_{%
\mathrm{odd}}\sim \varepsilon ^{3/4}.  \label{eps}
\end{equation}

Taking estimates (\ref{eps}) into account, the consideration of Eqs. (\ref%
{Ulimit}) and (\ref{Vlimit}) leads, first, to an approximate relation
between $U_{\mathrm{even}}$ and $U_{\mathrm{odd}}$:%
\begin{equation}
U_{\mathrm{odd}}=-\frac{1}{2}\frac{dU_{\mathrm{even}}}{dx}.  \label{oe}
\end{equation}%
After elimination of $U_{\mathrm{odd}}$ by means of Eq. (\ref{oe}), the
remaining equation for $U_{\mathrm{even}}(x)$ is%
\begin{equation}
\varepsilon U_{\mathrm{even}}=\frac{d^{2}U_{\mathrm{even}}}{dx^{2}}+2U_{%
\mathrm{even}}^{5}.  \label{quintic}
\end{equation}%
An obvious soliton solutions to Eq. (\ref{quintic}) is
\begin{equation}
U_{\mathrm{even}}^{\mathrm{(sol)}}(x)=\frac{\left( 3\varepsilon /2\right)
^{1/4}}{\sqrt{\cosh \left( 2\sqrt{\varepsilon }x\right) }},  \label{e-sol}
\end{equation}%
cf. Eq. (\ref{exact}). Then, the odd component of the soliton is produced by
the substitution of this in Eq. (\ref{oe}):%
\begin{equation}
U_{\mathrm{odd}}^{\mathrm{(sol)}}(x)=\frac{1}{2}\left( \frac{3}{2}%
\varepsilon ^{3}\right) ^{1/4}\frac{\sinh \left( 2\sqrt{\varepsilon }%
x\right) }{\left[ \cosh \left( 2\sqrt{\varepsilon }x\right) \right] ^{3/2}}.
\label{o-sol}
\end{equation}%
Expressions (\ref{e-sol}) and (\ref{o-sol}) agree with estimates (\ref{eps}%
), and the limit value of the total norm of the gap soliton coincides with
the TS value (\ref{Norm}):%
\begin{equation}
N\left( \varepsilon \rightarrow 0\right) \equiv 2\int_{-\infty }^{+\infty }
\left[ U_{\mathrm{even}}^{2}(x)+U_{\mathrm{odd}}^{2}(x)\right] _{\varepsilon
\rightarrow 0}dx=\sqrt{3/2}\pi \approx 3.85.  \label{TS2}
\end{equation}

\begin{figure}[tbp]
\centering{\subfigure[]{\includegraphics[scale=0.35]{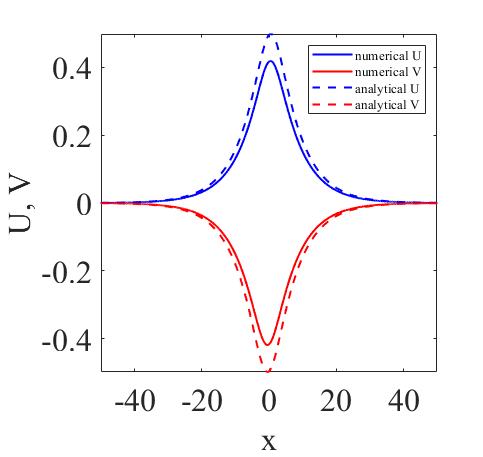}} %
\subfigure[]{\includegraphics[scale=0.35]{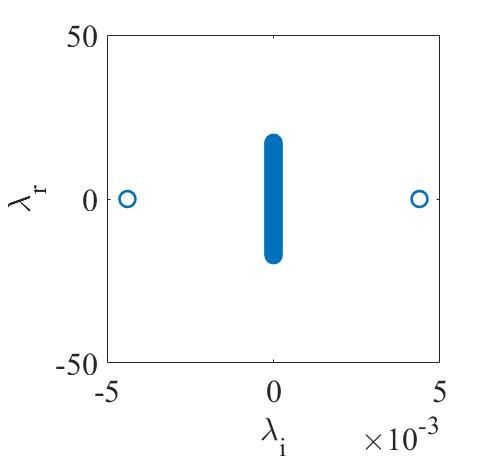}} \newline
\subfigure[]{\includegraphics[scale=0.33]{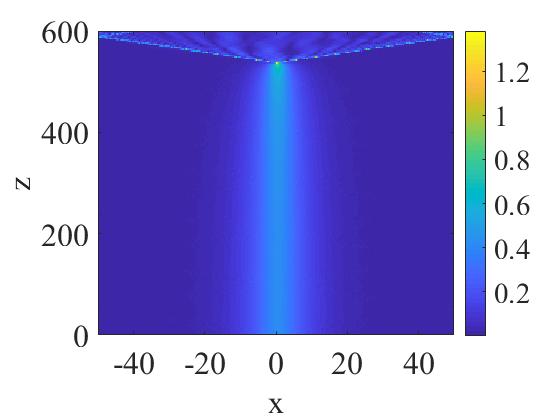}} \subfigure[]{%
\includegraphics[scale=0.33]{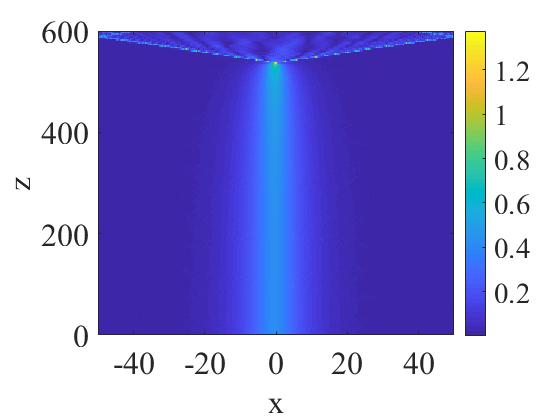}}}
\caption{{}(a) Numerically found profiles of components $U$ and $V$ of the
gap soliton, and their approximate analytical counterparts, given by Eqs. (%
\protect\ref{eo}), (\protect\ref{oe}), (\protect\ref{e-sol}) and (\protect
\ref{antiskew}), at $k=-0.99$ (i.e., $\protect\varepsilon =0.01$ in Eq. (%
\protect\ref{<<})). The total norm of the numerical solution is $N\approx
3.78$, to be compared to the TS value (\protect\ref{TS2}). (b) The spectrum
of stability eigenvalues for small perturbation modes around this soliton.
(c,d) Weakly unstable evolution of the soliton shown in (a). }
\label{fig10}
\end{figure}

The numerical solution for gap solitons near the bottom edge of the bandgap
produces a profile shown in Fig. \ref{fig10}, which is close to the
analytical approximation given by Eqs. (\ref{e-sol}) and (\ref{o-sol}). Note
that, even in the case of $\varepsilon =0.01$ displayed in the figure, the
soliton's amplitude is not really small.

The numerical solution of the equations produced by the linearization of
Eqs. (\ref{ugap}) and (\ref{vgap}) for small perturbations around the gap
solitons yields unstable eigenvalues. This conclusion complies with the
above-mentioned conjecture that skew-antisymmetric solitons are subject to
instability, all the gap solitons belonging to this type. In the case shown
in Fig. \ref{fig10}, the unstable eigenvalues are small and, accordingly,
direct simulations demonstrate that the perturbed gap soliton survives long
propagation, over the distance exceeding $10$ diffraction lengths, hence
these solitons are physically relevant states.

Deeper into the bandgap, the shape of the gap solitons becomes sharper, and
their instability gets stronger. In particular, at the midpoint of the
bandgap, $k=0$, the numerically found soliton's shape, displayed in Fig. \ref%
{fig11}, features strong spatial asymmetry mentioned above (see Eqs. (\ref%
{x>0}) and (\ref{x<0})). Its instability develops over the propagation
distance exceeding $\sim 5$ diffraction length. This numerical solution
corroborates the exact analytical results given by Eqs. (\ref{k=0}) and (\ref%
{x=0}).
\begin{figure}[tbp]
\centering{\subfigure[]{\includegraphics[scale=0.35]{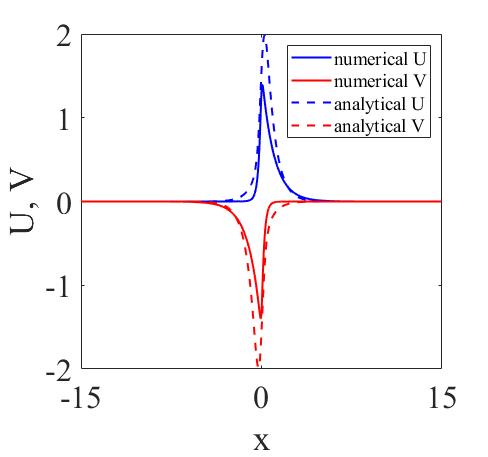}} %
\subfigure[]{\includegraphics[scale=0.35]{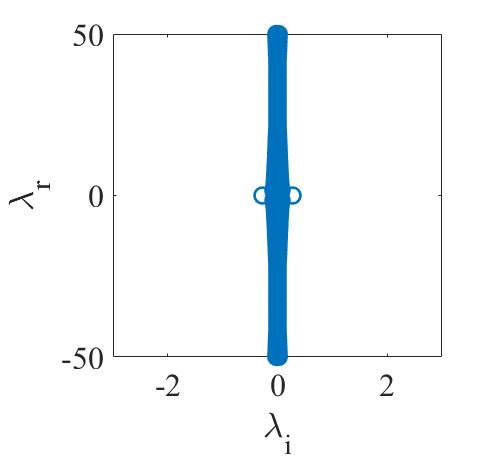}} \subfigure[]{%
\includegraphics[scale=0.23]{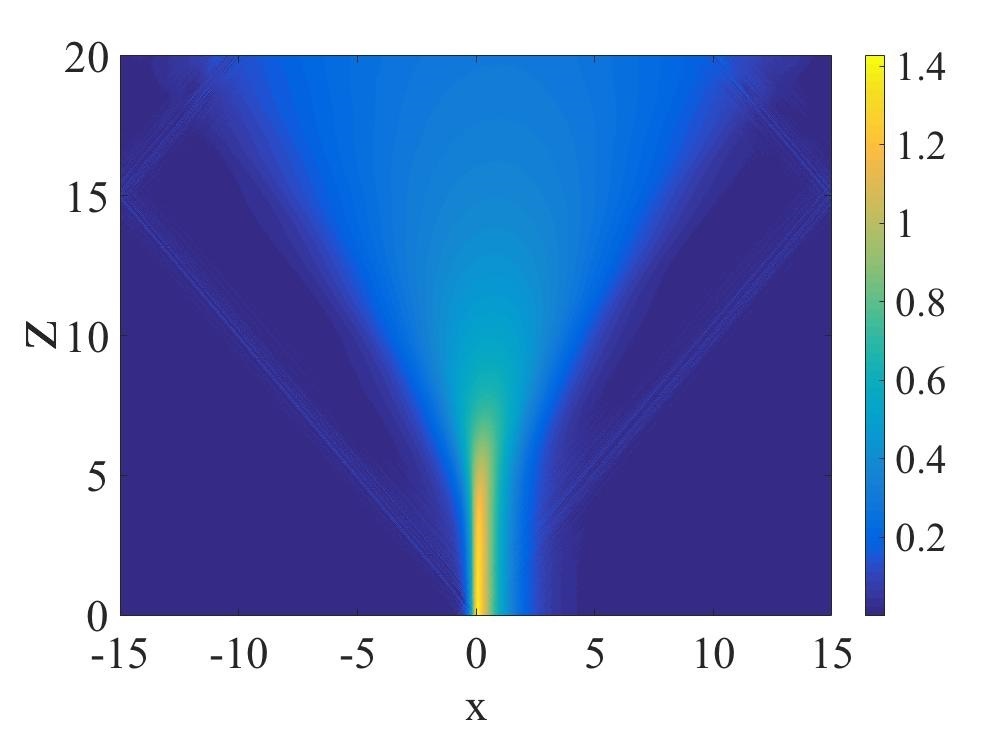}} \subfigure[]{%
\includegraphics[scale=0.23]{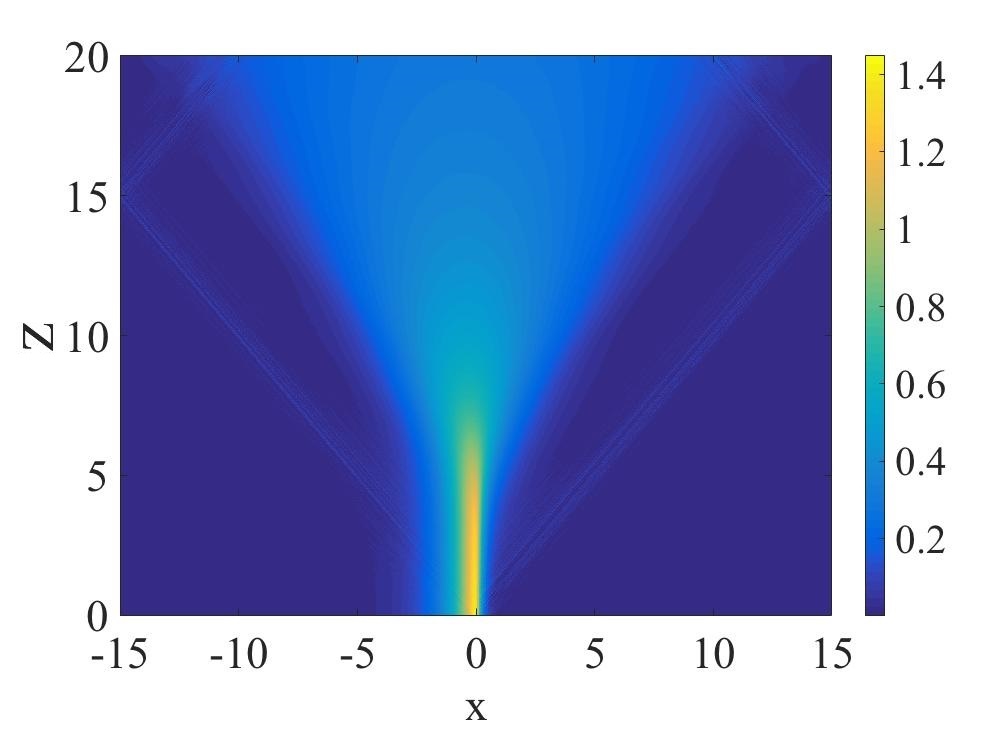}}}
\caption{{}The same as in Fig. \protect\ref{fig10}, but for the gap soliton
with $k=0$, taken at the midpoint of the bandgap. Values of the fields at $%
x=0$ are exactly predicted by Eq. (\protect\ref{U0}). The total norm of this
soliton is $N\approx 2.78$, which is essentially smaller than the TS value (%
\protect\ref{TS2}),}
\label{fig11}
\end{figure}

\begin{figure}[tbp]
\centering{\subfigure[]{\includegraphics[scale=0.35]{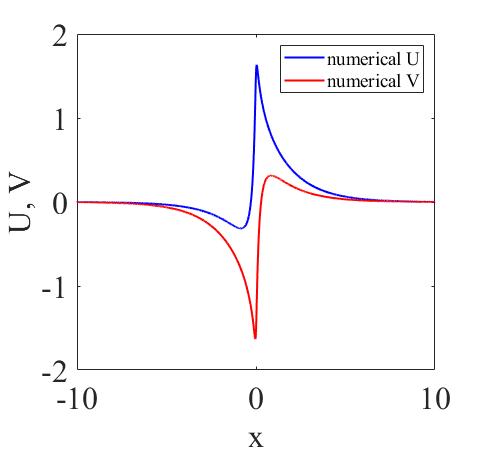}} %
\subfigure[]{\includegraphics[scale=0.35]{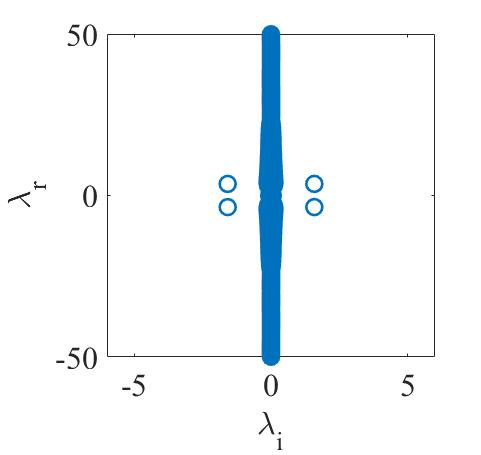}} \newline
\subfigure[]{\includegraphics[scale=0.2]{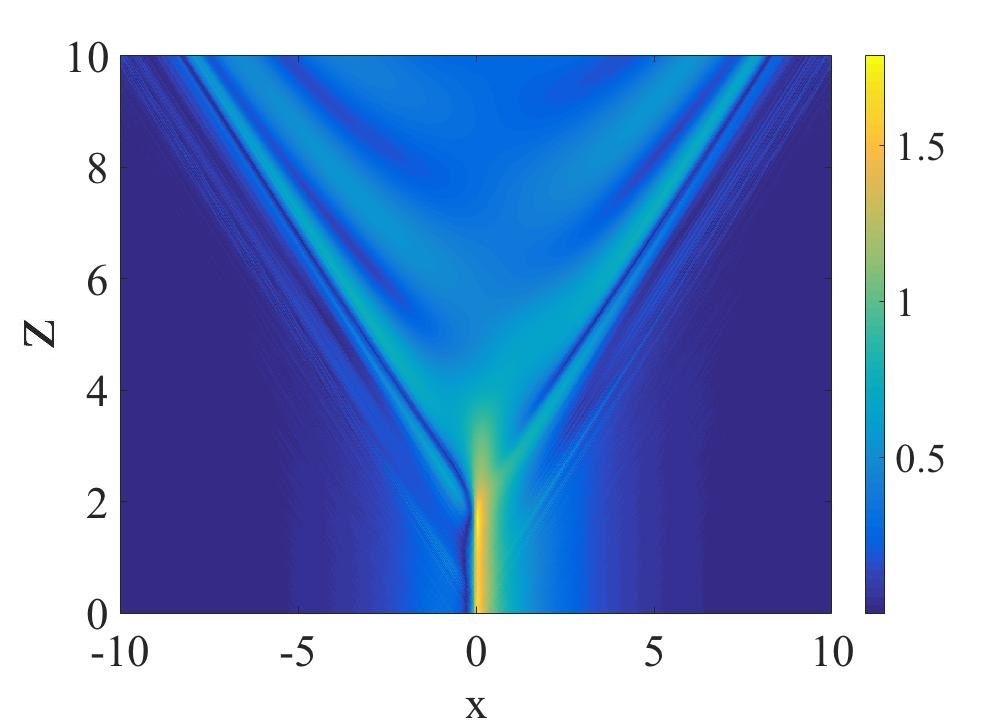}} \subfigure[]{%
\includegraphics[scale=0.15]{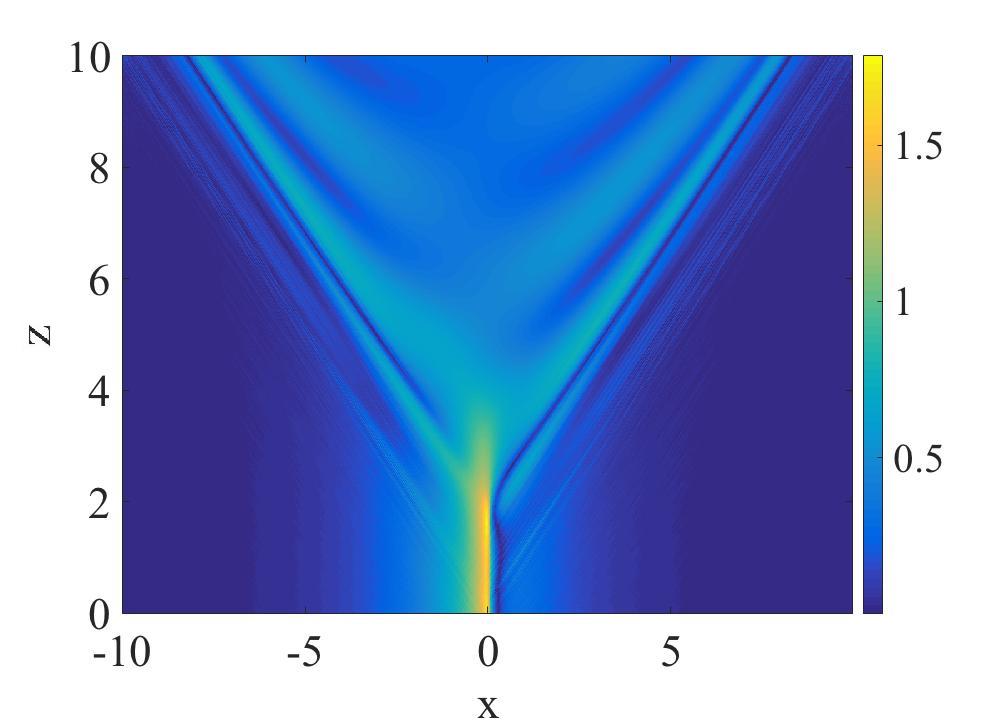}}}
\caption{{}The same as in Fig. \protect\ref{fig11}, but for the gap soliton
close to the top edge of the bandgap, at $k=+0.8$. Values of the fields at
the zero-crossing points and at $x=0$ are exactly predicted, severally, by
Eq. (\protect\ref{V=0}) and (\protect\ref{U0}). The total norm of the
soliton is $N\approx 3.94$, which exceeds the TS value (\protect\ref{TS2}).}
\label{fig12}
\end{figure}

The numerical solution produces gap solitons as well at $k>0$, as shown in
Fig. \ref{fig12}. Their shape features zero crossings, in exact agreement
with Eq. (\ref{V=0}). In this case, the instability is strong, leading to
destruction of the soliton after passing $\simeq 1$ diffraction length.

\begin{figure}[tbp]
\centering{\includegraphics[scale=0.45]{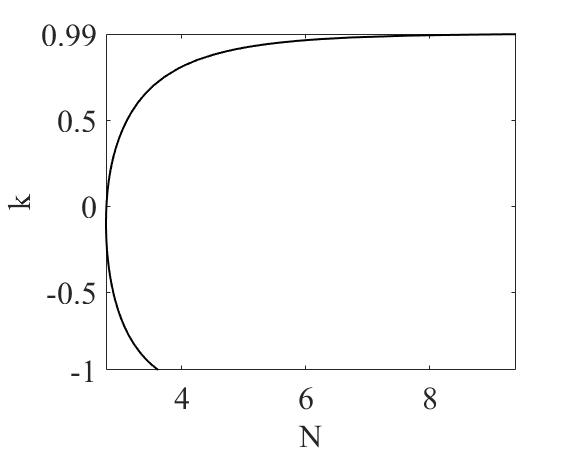}}
\caption{{}The numerically found dependence $k(N)$ for the gap-solitons
family, produced by the simplified diffractionless system (\protect\ref{ugap}%
), (\protect\ref{vgap}). }
\label{fig13}
\end{figure}

Finally, the gap-soliton family as a whole is characterized by the $k(N)$
dependence, which is shown in Fig. \ref{fig13}, as obtained from the
numerical solution. It is worthy to note the difference of this double-value
dependence from its monotonous single-value counterpart obtained in the full
system, which includes the diffraction terms (second derivatives), cf. Fig. %
\ref{fig8}(d). The limit value of $N$ corresponding to the bottom of the
bandgap, $k=-1$, is given by Eq. (\ref{TS2}).\textbf{\Large \ }\newline

Because the value of the solution with $k=1$ at $x=0$, as given by Eq. (\ref%
{U0}), is finite, $U(x=0,k=1)=-V\left( x=0,k=1\right) =3\cdot 2^{1/4}$, and
the width of the soliton diverges $\sim \left( 1-k^{2}\right) ^{-1/2}$ in
the limit of $k\rightarrow 1$, as per Eqs. (\ref{tails}), the integral norm
diverges too in this limit.

\section{Conclusion}

The objective of this work is to expand the recently proposed mechanism for
stabilizing TSs (Townes solitons) by means of the linear SOC
(spin-orbit-coupling) terms in binary BEC, or ones emulating SOC in other
physical settings. In Refs. \cite{BLi} and \cite{Barcelona}, this mechanism
was elaborated for 2D matter-wave solitons realizing SOC in BEC, and for 2D
spatiotemporal optical solitons in dual-core planar waveguides. In both
cases, TSs existing in the absence of SOC or quasi-SOC are the usual 2D
solitons created in the unstable form by the cubic focusing nonlinearity.
Here, we elaborate the mechanism for the stabilization of the 1D variety of
TSs, in a two-component system with the quintic self-focusing. This setting
may be realized for spatial solitons in a dual-core planar optical
waveguides dominated by the quintic nonlinearity, with the SOC-emulating
terms represented by skewness in the tunnel coupling between parallel cores
of the coupler. The results identify a vast stability region for
skew-symmetric two-component solitons, in the main (semi-infinite) and annex
(finite) bandgaps alike. Thus, the SOC-driven stabilization method is a
universal one, being applicable to TS in both 2D and 1D settings, with the
cubic and quintic nonlinearities, respectively. As a part of the analysis,
we have also considered asymmetric solitons in the quintic coupler in the
absence of SOC.

An extension of the analysis, also elaborated in this work, is its
application to broad solitons, for which the usual diffraction terms are
negligible, and the system is simplified to one with a finite bandgap. In
this case, some results for the solitons populating the gap can be obtained
in an exact analytical form, and a subfamily of the gap solitons near the
bottom of the gap is constructed in an approximate analytical form. On the
contrary to the full system, the reduced diffractionless one maintains only
skew-antisymmetric solitons, which are unstable, although the instability is
weak for the solitons constructed near the bandgap's bottom.

The present analysis can be further developed in several directions. In
particular, a challenging possibility is to develop a systematic analysis of
tilted solitons, in the full and reduced systems alike. Other relevant
issues are interactions of solitons in these systems, as well as effects of
dissipation.

\begin{acknowledgments}
This work was supported, in part, by the Israel Science Foundation through
grant No. 1287/17. Z.C. acknowledges an excellence scholarship provided by
the Tel Aviv University.
\end{acknowledgments}


\begin{thebibliography}{99}
\bibitem{Berge} Berg\'{e} L. Wave collapse in physics: Principles and
applications to light and plasma waves. Phys. Rep. 1998;303:259-372.

\bibitem{Fibich} Fibich G. The Nonlinear Schr\"{o}dinger Equation: Singular
Solutions and Optical Collapse. Springer: Heidelberg, 2015.

\bibitem{Townes} Chiao RY, Garmire E, Townes CH, Self-trapping of optical
beams. Phys. Rev. Lett. 1964;13:479-482.

\bibitem{1DT1} Abdullaev FK, Salerno M. Gap-Townes solitons and localized
excitations in low-dimensional Bose-Einstein condensates in optical
lattices. Phys. Rev. A 2005;72:033617.

\bibitem{1DT2} Senthilnathan K, Li Q, Nakkeeran K, Wai PKA. Robust
pedestal-free pulse compression in cubic-quintic nonlinear media. Phys. Rev.
A 2008;78:033835.

\bibitem{KA} Kivshar YS, Agrawal GP. Optical Solitons: From Fibers to
Photonic Crystals. San Diego: Academic Press; 2003.

\bibitem{Robinson} Robinson PA. Nonlinear wave collapse and strong
turbulence. Rev. Mod. Phys. 1997;69:507-573.

\bibitem{Pit} Pitaevskii P, Stringari S, Bose-Einstein Condensation, Oxford
University Press: Oxford, 2003.

\bibitem{2005} Malomed BA, Mihalache D, Wise F, Torner L. Spatiotemporal
optical solitons. J. Optics B 2005;7:R53--R72.

\bibitem{2008} Mihalache D. Three-dimensional dissipative optical solitons.
Cent. Eur. J. Phys. 2008;6:582-587.

\bibitem{viewpoint} Malomed BA, Mihalache D, Wise F, Torner L. Viewpoint: On
multidimensional solitons and their legacy in contemporary atomic, molecular
and optical physics. J. Phys. B: At. Mol. Opt. Phys. 2016;49:170502.

\bibitem{2016} Malomed BA. Multidimensional solitons: Well established
results and novel findings. Eur. Phys. J. Special Top. 2016;225:2507-2532.

\bibitem{2019} Kartashov Y, Astrakharchik G, Malomed B, Torner L. Frontiers
in multidimensional self-trapping of nonlinear fields and matter. Nature
Reviews Physics 2019;1:185-197; https://doi.org/10.1038/s42254-019-0025-7.

\bibitem{Michinel} Quiroga-Teixeiro M, Michinel H. Stable azimuthal
stationary state in quintic nonlinear optical media. J. Opt. Soc. Amer. B
1997;14:2004-2009.

\bibitem{Cid1} Falc\~{a}o-Filho, de Ara\'{u}jo CB, Boudebs G, Leblond H,
Skarka V. Robust two-dimensional spatial solitons in liquid carbon
disulfide. Phys. Rev. Lett. 2013;110;013901.

\bibitem{Cid2} Reyna AS, de Ara\'{u}jo CB. High-order optical nonlinearities
in plasmonic nanocomposites -- a review. Adv. Opt. Phot. 2017;9:720-774.

\bibitem{BBB} Baizakov BB, Malomed BA, Salerno M. Multidimensional solitons
in periodic potentials. Europhys. Lett. 2003;63:642-648.

\bibitem{Ziad} Yang J, Musslimani ZH. Fundamental and vortex solitons in a
two-dimensional optical lattice. Opt. Lett. 2003;28:2094-2096.

\bibitem{Petrov} Petrov DS. Quantum mechanical stabilization of a collapsing
Bose-Bose mixture. Phys. Rev. Lett. 2015;115:155302.

\bibitem{PA} Petrov DS, Astrakharchik GE. Ultradilute low-dimensional
liquids. Phys. Rev. Lett. 2016;117:100401.

\bibitem{Polish} \.{Z}in P, Pylak M, Wasak T, Gajda M, Idziaszek Z. Quantum
Bose-Bose droplets at a dimensional crossover, Phys. Rev. A
2018;98;051603(R).

\bibitem{Santos} Ilg T, Kumlin J, Santos L, Petrov DS, Buchler HP.
Dimensional crossover for the beyond-mean-field correction in Bose gases.
Phys. Rev. A 2018;98; 051604.

\bibitem{LHY} Lee TD, Huang K, Yang CN. Eigenvalues and eigenfunctions of a
Bose system of hard spheres and its low temperature properties. Phys. Rev.
1957; 106: 1135-1145.

\bibitem{Leticia1} Cabrera C, Tanzi L, Sanz J, Naylor B, Thomas P, Cheiney
P, Tarruell L, Quantum liquid droplets in a mixture of Bose-Einstein
condensates. Science 2018;359:301-304.

\bibitem{Leticia2} Cheiney P, Cabrera CR, Sanz J, Naylor B, Tanzi L,
Tarruell L. Bright soliton to quantum droplet transition in a mixture of
Bose-Einstein condensates. Phys. Rev. Lett. 2018;120:135301.

\bibitem{Inguscio1} G. Semeghini, G. Ferioli, L. Masi, C. Mazzinghi, L.
Wolswijk, F. Minardi, M. Modugno, G. Modugno, M. Inguscio, M. Fattori.
Self-bound quantum droplets of atomic mixtures in free space. Phys. Rev.
Lett. 2018;120:235301.

\bibitem{Inguscio2} Ferioli G, Semeghini G, Masi L, Giusti G, Modugno G,
Inguscio M, Gallemi A, Recati A, Fattori M. Collisions of self-bound quantum
droplets. Phys. Rev. Lett. 2019;122:090401.

\bibitem{hetero} C. D'Errico, A. Burchianti, M. Prevedelli, L. Salasnich, F.
Ancilotto, M. Modugno, F. Minardi, C. Fort. Observation of quantum droplets
in a heteronuclear bosonic mixture. Phys. Rev. Research 2019;1:033155.

\bibitem{Barcelona} Kartashov YV, Malomed BA, Tarruell L, Torner L.
Three-dimensional droplets of swirling superfluids. Phys Rev. A
2018;98:013612.

\bibitem{Raymond} Li Y, Chen Z, Luo Z, Huang C, Tan H, Pang W, Malomed BA.
Two-dimensional vortex quantum droplets. Phys. Rev. A 2018;98:063602.

\bibitem{Pfau1} H. Kadau, M. Schmitt, M. Wentzel, C. Wink, T. Maier, I.
Ferrier-Barbut, and T. Pfau, Observing the Rosenzweig instability of a
quantum ferrofluid. Nature 2016;530;194-197.

\bibitem{Pfau2} M. Schmitt, M. Wenzel, 491 F. Bottcher, I. Ferrier-Barbut
and T. Pfau. Self-bound droplets of a dilute magnetic quantum liquid. Nature
2016;539;259-262.

\bibitem{Pfau3} Wachtler F, Santos L. Ground-state properties and elementary
excitations of quantum droplets in dipolar Bose-Einstein condensates. Phys.
Rev. A 2016;94:043618.

\bibitem{Pfau4} Baillie D, Blakie PB. Droplet crystal ground states of a
dipolar Bose gas. Phys. Rev. Lett. 2018;121;195301.

\bibitem{Pfau5} Cidrim A, dos Santos FEA, Henn EAL, Macr\'{\i} T. Vortices
in self-bound dipolar droplets. Phys. Rev. A 2018;98;023618.

\bibitem{Spielman} Lin YJ, Jimenez-Garcia K, Spielman IB. Spin-orbit-coupled
Bose--Einstein condensates. Nature 2011;471:83-86.

\bibitem{Goldman} Goldman N, Juzeli\={u}nas G, \"{O}hberg P, Spielman IB.
Light-induced gauge fields for ultracold atoms. Rep. Progr. Phys.
2014;77:126401.

\bibitem{Zhai} Zhai H. Degenerate quantum gases with spin-orbit coupling: a
review. Rep. Progr. Phys. 2015;78:026001.

\bibitem{super1} Adams EN, Blount EI. Energy bands in the presence of an
external force field-II. Anomalous velocities. J. Phys. Chem. Solids
1959;10:286-303.

\bibitem{super2} Mardonov Sh, Sherman EYa, Muga JG, Wang H-W, Ban Y, Chen X.
Collapse of spin-orbit-coupled Bose-Einstein condensates. Phys. Rev. A
2015;91:043604.

\bibitem{XiChen} Li J., Malomed BA, Li W., Chen X., Sherman EYa. Coupled
density-spin Bose-Einstein condensates dynamics and collapse in systems with
quintic nonlinearity. Communications in Nonlinear Science and Numerical
Simulation. 2020;82:105045.

\bibitem{BLi} Sakaguchi H, Li B, Malomed BA. Creation of two-dimensional
composite solitons in spin-orbit-coupled self-attractive Bose-Einstein
condensates in free space. Phys. Rev. E 2014;89:032920.

\bibitem{Evgeny} Sakaguchi H, Sherman EYa, Malomed BA. Vortex solitons in
two-dimensional spin-orbit coupled Bose-Einstein condensates: Effects of the
Rashba-Dresselhaus coupling and the Zeeman splitting. Phys. Rev. E
2016;94:032202.

\bibitem{quasi1D} Kartashov YV, Torner L, Modugno M, Sherman EYa, Malomed
BA, and Konotop VV. Multidimensional hybrid Bose-Einstein condensates
stabi-lized by lower-dimensional spin-orbit coupling. Phys. Rev. Research
2020;2:013036.

\bibitem{HPu} Zhang YC, Zhou ZW, Malomed BA, Pu H. Stable solitons in three
dimensional free space without the ground state: Self-trapped Bose-Einstein
condensates with spin-orbit coupling. Phys. Rev. Lett. 2015;115:253902.

\bibitem{emulation} Kartashov YV, Malomed BA, Konotop VV, Lobanov VE, Torner
L. Stabilization of solitons in bulk Kerr media by dispersive coupling. Opt.
Lett. 2015;40:1045-1048.

\bibitem{NJP} Sakaguchi H, Malomed BA. One- and two-dimensional solitons in $%
\mathcal{PT}$-symmetric systems emulating spin-orbit coupling. New J. Phys.
2016;18:105005.

\bibitem{Lior} Albuch L, Malomed B A. Transitions between symmetric and
asymmetric solitons in dual-core systems with cubic-quintic nonlinearity.
Mathematics and Computers in Simulation 2007;74:312-322.

\bibitem{Wabnitz} Wright EM, Stegeman GI, Wabnitz S. Solitary-wave decay and
symmetry-breaking instabilities in two-mode fibers. Phys. Rev. A
1989;40:4455.

\bibitem{Pare} Par\'{e} C, Florja\'{n}czyk M. Approximate model of soliton
dynamics in all-optical fibers. Phys. Rev. A 1990;41:6287-6295.

\bibitem{Maimistov} Maimistov AI. Propagation of a light pulse in nonlinear
tunnel-coupled optical waveguides. Sov. J Quantum Electron 1991;21:687-690.

\bibitem{Wabnitz2} Romagnoli M, Trillo S, Wabnitz S. Soliton switching in
nonlinear couplers. Opt Quantum Electron 1992;24:1237-1267.

\bibitem{Skinner} Chu PL, Malomed BA, Peng GD, Skinner I. Soliton dynamics
in periodically modulated directional couplers. Phys. Rev. E
1994;49:5763-5767.

\bibitem{Peng} Malomed BA. Solitons and nonlinear dynamics in dual-core
optical fibers. In: Handbook of Optical Fibers (Peng GD, editor), Springer;
2018.

\bibitem{Landau} Landau LD, Lifshitz EM. Quantum Mechanics. Nauka
Publishers: Moscow, 1974.

\bibitem{Thawatchai} Mayteevarunyoo T, Malomed BA, Dong G. Spontaneous
symmetry breaking in a nonlinear double-well structure. Phys. Rev. A
2008;78:053601.

\bibitem{VK} Vakhitov NG, Kolokolov AA. Stationary solutions of the wave
equation in a medium with nonlinearity saturation. Radiophys. Quantum
Electron. 1973;16:783-789.

\bibitem{Yang} Yang J. Nonlinear Waves in Integrable and Nonintegrable
Systems SIAM: Philadelphia, 2010.

\bibitem{skewsymm} Mayteevarunyoo T, Malomed BA. Skew-symmetric vortices and
solitons in crossed-lattice potentials. J. Opt. A: Pure Appl. Opt.
2009;11:094015.

\bibitem{cross} Fan Z, Shi Y, Liu Y, Pang W, Li Y, Malomed BA.
Cross-symmetric dipolar-matter-wave solitons in double-well chains. Phys.
Rev. E 2017;95:032226.

\bibitem{GZ} Li Y, Liu Y, Fan Z, Pang W, Fu S, Malomed BA. Two-dimensional
dipolar gap solitons in free space with spin-orbit coupling. Phys. Rev. A
2017;95:063613.

\bibitem{Fukuoka} Sakaguchi H, Malomed BA. One- and two-dimensional gap
solitons in spin-orbit-coupled systems with Zeeman splitting. Phys. Rev. A
2018;97:013607.

\bibitem{gapsol1} Chen W, Mills DL. Gap solitons and the nonlinear optical
response of superlattices. Phys. Rev. Lett. 1987;58:160-163.

\bibitem{gapsol2} de Sterke CM, Sipe JE. Gap solitons. Progr. Optics
1994;33:203-260.
\end{thebibliography}
\end{document}